\documentclass[twocolumn,twocolappendix]{aastex701}

\usepackage{amsmath}
\usepackage{amssymb}
\usepackage{breqn}
\usepackage{enumerate}
\usepackage{isotope}
\usepackage{hyperref}

\newcommand{\be}{\begin{eqnarray}}
\newcommand{\ee}{\end{eqnarray}}
\newcommand{\rsun}{\ensuremath{R_{\odot}}}
\newcommand{\msun}{\ensuremath{M_{\odot}}}

\newcommand{\grad}{\ensuremath{\boldsymbol{\nabla}}}
\newcommand{\vel}{\ensuremath{\boldsymbol{v}}}
\newcommand{\mom}{\ensuremath{\boldsymbol{p}}}
\newcommand{\ddt}[1]{\ensuremath{\frac{\partial #1}{\partial t}}}

\newcommand{\mycomment}[1]{}

\newcommand{\lr}[1]{\left( #1 \right )}
\newcommand{\lrb}[1]{\left[ #1\right]}

\newcommand{\thermoD}[3]{\left(\frac{\partial #1}{\partial #2}\right)_{#3}}
\newcommand{\RNum}[1]{\uppercase\expandafter{\romannumeral #1\relax}}
\newcommand{\Eej}{\ensuremath{E_{\rm ej}}}
\newcommand{\Mej}{\ensuremath{M_{\rm ej}}}

\newcommand{\change}[1]{\textbf{#1}}
\renewcommand{\change}[1]{#1}

\begin{document}

\title{The First Day of a Type Ia Supernova from a Double-Degenerate Binary}


\correspondingauthor{Gabriel Kumar}
\email{gabrielkumar@hotmail.com}

\author[0009-0004-0856-0915]{Gabriel Kumar}
\affiliation{College of Creative Studies, University of California, Santa Barbara, CA 93106, USA}
\email{gabrielkumar@hotmail.com}

\author[0000-0002-7174-8273]{Logan J. Prust}
\affiliation{Kavli Institute for Theoretical Physics, University of California, Santa Barbara, CA 93106, USA}
\email{ljprust@kitp.ucsb.edu}

\author[0000-0001-8038-6836]{Lars Bildsten}
\affiliation{Kavli Institute for Theoretical Physics, University of California, Santa Barbara, CA 93106, USA}
\affiliation{Department of Physics, University of California, Santa Barbara, CA 93106, USA}
\email{bildsten@kitp.ucsb.edu}

\begin{abstract}
Supernovae in binary star systems involve a hydrodynamical interaction between the ejecta and a binary companion. This collision results in shock heating and a modified density structure for the ejecta, both of which affect the light curve. As highlighted by Kasen, these considerations are particularly relevant for type Ia supernovae, as the companion is expected to be Roche-lobe filling at the time of the explosion. We simulate here the interaction between type Ia supernova ejecta and a white dwarf donor using Athena++, finding the formation of a low-density wake extending to higher velocities than the unperturbed ejecta. Radiation hydrodynamics is then used to generate synthetic light curves for the first day after the explosion for a range of viewing angles. We find that the hot, high-velocity, shocked ejecta yields $L>10^{40}$ ergs/s over half the sky in the first few hours. The photosphere within the shock-heated ejecta cools and recedes in velocity space, partially obscuring it from view, as heating from radioactive nickel becomes increasingly important in driving the supernova's luminosity. By one day after the explosion, the luminosity measured by observers looking directly into the wake is dimmer than that of a normal type Ia supernova by 15 percent due to the modified density structure.


\end{abstract}


\section{Introduction} \label{sec:intro}

Type Ia supernovae (SNIa) can be triggered by accretion onto a white dwarf (WD) from a binary companion or from the merger of two white dwarfs. In the former case, the donor may be either degenerate or non-degenerate \citep[for a review see][]{2025A&ARv..33....1R}. Regardless, the collision between the ejecta and donor may have lasting effects both on the subsequent evolution of the donor and on the morphology of the ejecta. The effects on the donor---such as mass stripping, pollution, and entropy deposition---have been explored in a variety of hydrodynamical simulations \citep{2000ApJS..128..615M,2015A&A...584A..11L,2015MNRAS.449..942P,2018ApJ...868...90T,2019ApJ...887...68B, Wong_2024}. Though these studies have considered both the single- and double-degenerate scenarios, the fact that mass transfer requires the donor to fill its Roche lobe means that the solid angle subtended by the donor relative to the accretor is independent of the degeneracy of the donor, and approaches $10\%$ of the whole sky. \change{In the single-degenerate case, the donor may be farther from the explosion if the accretor gains sufficient angular momentum to support a super-Chandrasekhar mass \citep{2011ApJ...730L..34J, 2011ApJ...738L...1D, 2012ApJ...756L...4H}.} The encounter of the expanding ejecta with the donor results in a wake that is nearly a factor of two larger. 

In the double-degenerate case, the supernova may be achieved by the detonation of a thin shell of accreted helium surrounding a carbon-oxygen (CO) WD \citep{2021ApJ...919..126B}, driving a converging shock wave into the accretor which then triggers a second detonation \citep{2014ApJ...785...61S}. This formation channel necessitates a small orbital separation at the time of detonation, with an orbital velocity of $\approx$800--1500  km/s for the donor. The population of hypervelocity white dwarfs observed by Gaia \citep{2018ApJ...865...15S, 2023OJAp....6E..28E} and SDSS \citep{2025MNRAS.tmp..954H} are likely to be the surviving donors of double-degenerate SNIa. Furthermore, the deposition of entropy into the interior of the donor places it in an unusual location on the Hertzsprung-Russell diagram \citep{2019ApJ...887...68B, Wong_2024, 2019ApJ...872...29Z}.

Our focus here is on the thermodynamics and subsequent radiation from the conical wake in the ejecta created by the donor interaction. Filled with low density, high temperature gas due to heating from the bow shock adorning the donor, this wake is a permanent feature of the ejecta. \citet{2010ApJ...708.1025K} showed that the wake has observational consequences for the early emission from the supernova, which has been substantiated in several cases: \citet{2016ApJ...820...92M} concluded that a blue bump in SN 2012cg was due to interaction with a non-degenerate donor, \citet{2022NatAs...6..568N} observed a plateau in the B-band between 1.0 and 12.4 hours after first light in SN 2018aoz, and \citet{2015MNRAS.454.1192L} showed that a UV flash in SN Ia iPTF14atg was due to interaction with a non-degenerate donor. 
On the other hand, \citet{2016ApJ...826...96P} and \citet{2017MNRAS.472.2787N} both note that an excess of $^{56}$Ni in the outermost ejecta can masquerade as a collision with the circumstellar medium (CSM) or a donor, and \citet{2023MNRAS.525..246H} show that some (but not all) SNIa multi-band light curves can be explained by CSM interaction. 

\citet{2010ApJ...708.1025K} performed the first calculation for the single-degenerate case.
Multiple groups have since studied the light curves and spectra of double-degenerate SNIa in various scenarios \citep{2021ApJ...922...68S,2024MNRAS.tmp.1883P,2024ApJ...972..200B} but focused on $t\gtrsim 5$ days after shock breakout, so it remains unclear to what degree this affects the synthetic observables at $t\leq 1$ day.
The double-degenerate case requires a special consideration of the equation of state of the ejecta. It is initially gas pressure-dominated, but becomes radiation-pressure dominated due to the bow shock. To calculate the resulting entropy of the post-shock material, a full equation of state spanning both regimes must be used. 
To this end, we perform hydrodynamical simulations of the donor interaction with SNIa ejecta to obtain the resulting ejecta structure. We then homologously evolve this structure and apply radiation hydrodynamics to predict the supernova luminosity from all parts of the sky in the first day after the explosion. 

We organize this paper as follows. The setup for our hydrodynamical simulations is described in section \ref{sec:setup}, where we show the results of these calculations. Subsequent homologous evolution of the ejecta (including $^{56}$Ni heating) is discussed in section \ref{sec:extrapolation}. Radiation hydrodynamics is used to obtain bolometric light curves of these events in section \ref{sec:radiationResults}. We discuss these results and explore the potential for future work in section \ref{sec:discussion}.

\section{Hydrodynamical Interaction and the Wake Structure} \label{sec:setup}

We perform hydrodynamical simulations using Athena++ \citep{athena++}, which solves the Euler equations
\be
\ddt{\rho} + \grad\cdot\rho\vel &=& 0, \label{eq:continuity}\\
\ddt{\rho\vel} + \grad\cdot(\rho\vel\vel^{T} + P \mathbb{I}) &=& 0, \label{eq:momentum}\\
\ddt{u} + \grad\cdot\left(e + P\right)\vel &=& 0 \label{eq:energy}
\ee
for material with density $\rho$, fluid velocity $\vel$, pressure $P$, energy density $e=u+\rho v^{2}/2$, and internal energy $u$, where $\mathbb{I}$ is the identity matrix. \change{A Zenodo record containing the Athena++ setup for all simulations reported in this paper as well as our analysis scripts can be found in \citet{kumar_2025_16388748}}.

\begin{figure}
    \includegraphics[width=\linewidth]{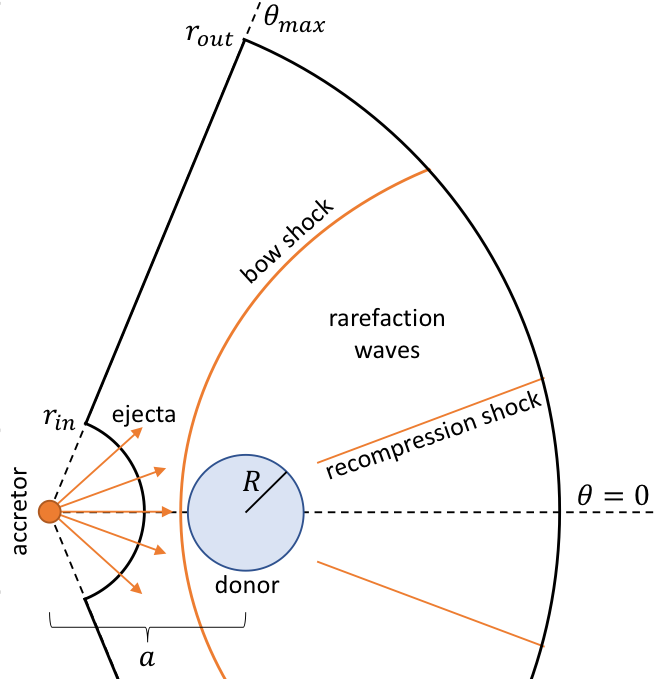}
    \caption{Diagram of our numerical setup highlighting several features of the resulting wake.}
    \label{fig:diagram}
\end{figure}

Most of the ejecta is initially gas-pressure dominated, but becomes radiation-pressure dominated when shocked or later heated by radioactive decay. We treat the gas as ideal and include radiation pressure assuming local thermodynamic equilibrium, so that the internal energy and pressure are
\be
u&=&a_rT^4+\frac{3}{2}\frac{\rho k_B T}{\mu m_p} \label{energyDensity},\\
P&=&\frac{1}{3}a_rT^4+\frac{\rho k_B T}{\mu m_p} \label{pressure},
\ee
where $T$ is the gas and radiation temperature, $a_r$ is the radiation constant, $k_B$ is the Boltzmann constant, $m_p$ is the proton mass, and $\mu$ is the mean molecular weight of the fully ionized plasma which we set to $4/3$ (fully ionized He). Closure of the Euler equations requires conversion between pressure and energy, and thus a determination of $T$. We accomplish this via a quartic solver which performs a root find to solve equations of the form $T^4+BT-A=0$. We also require the adiabatic sound speed $c_s^2=\Gamma_1 P/\rho$. Here $\Gamma_1$ is given by
\be
\Gamma_1=\thermoD{P}{\rho}{S}=\frac{32-24\beta-3\beta^2}{24-21\beta} \label{gamma}\
\ee
\citep{osti_5734165}, where $\beta$ is the ratio of gas pressure to total pressure.

We use a 3-D spherical-polar mesh with the explosion at the origin. The domain extends from an inner radial boundary $r_{\rm in}=0.13~\rsun$---where we inject the SN ejecta---to an outer boundary at $r_{\rm out}=10.8~\rsun$ where the gas freely outflows (Fig.~\ref{fig:diagram}). This $r_{\rm out}$ is sufficiently large to ensure that the gas exiting this boundary has reached homology, which we demonstrate below. The donor is located on the $\theta=0$ pole at an orbital separation of $a=0.27~\rsun$ with radius $R=0.08~\rsun$ and mass $M_{d}=0.344~\msun$, which is identical to Model 1 of \citet{2019ApJ...887...68B}. The domain extends out to a maximum polar angle of $80^{\circ}$, while the azimuthal angle $\phi$ spans the full range of $0$ to $2\pi$. The donor is treated as a reflective spherical boundary. We use the Gaussian ejecta profile from \citet{Wong_2024} with explosion energy $\Eej=0.97\times 10^{51}$ ergs and ejecta mass $\Mej=0.89~\msun$. More detailed information on the numerical setup is available in \citet{2025ApJ...982...60P}.

\begin{figure}[t!]
  \centering
  \begin{minipage}[b]{0.48\textwidth}
    \includegraphics[width=\textwidth]{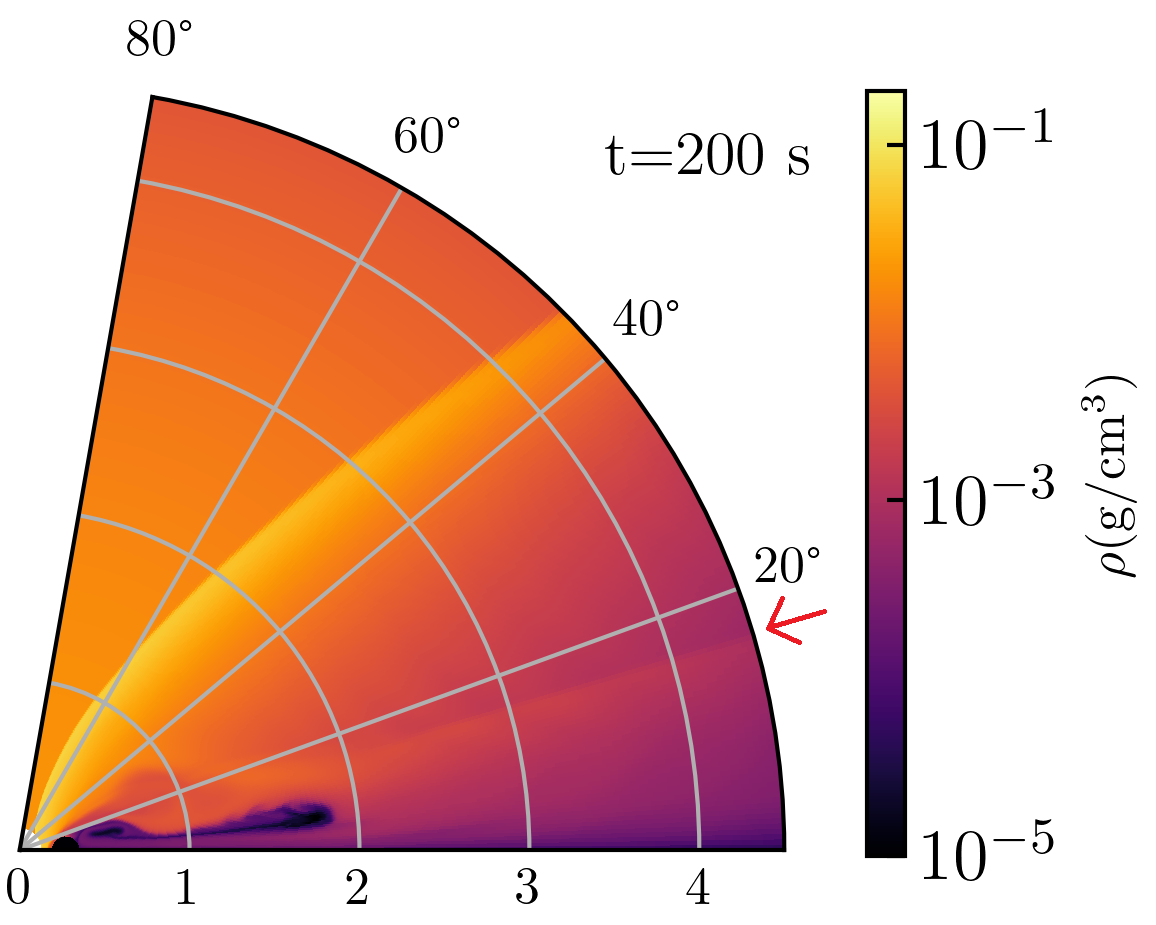}
  \end{minipage}
  \hfill
  \begin{minipage}[b]{0.48\textwidth}
    \includegraphics[width=\textwidth]{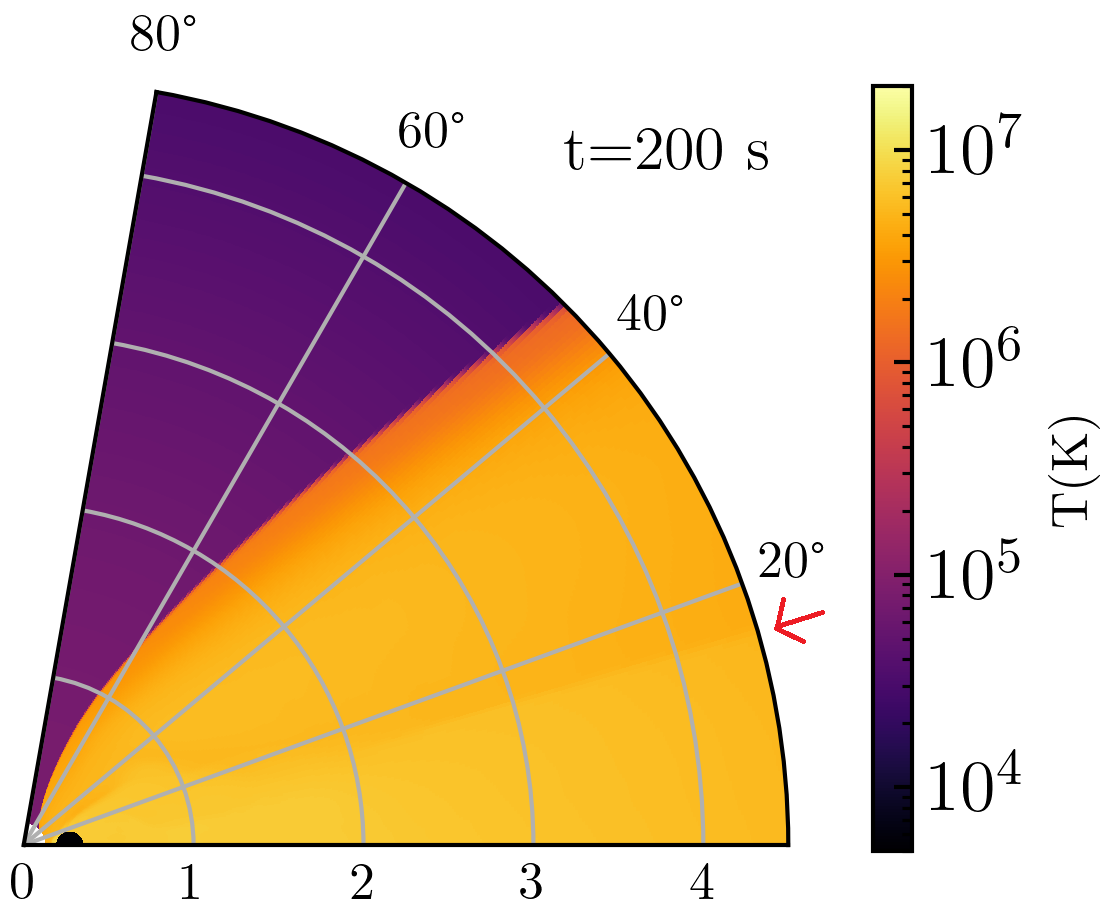}
  \end{minipage}
  \caption{Slice plots showing density (top) and temperature (bottom) on the $\phi=0$ plane 200 seconds after the supernova. The red arrows indicate the location of the secondary shock. }
  \label{fig:slicePlots}
\end{figure}

\begin{figure*}[h]
  \centering
  \begin{minipage}[b]{0.48\textwidth}
    \includegraphics[width=\textwidth]{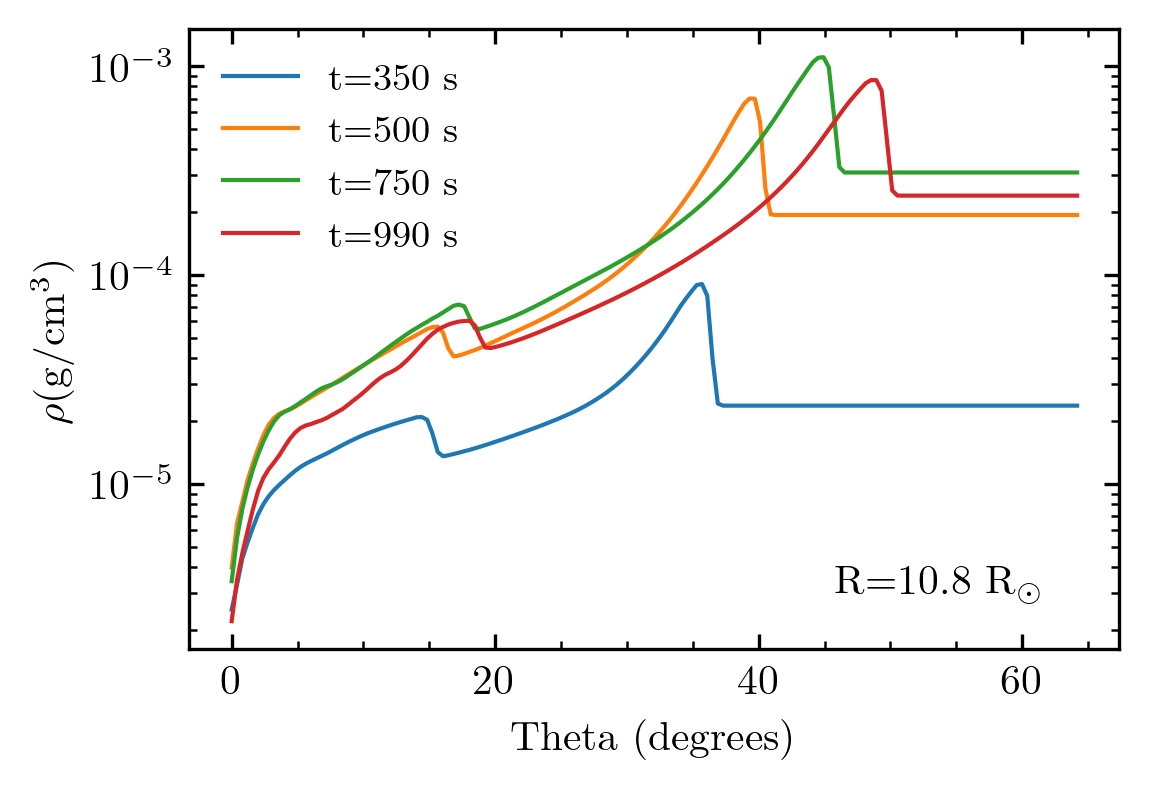}
  \end{minipage}
  \hfill
  \begin{minipage}[b]{0.48\textwidth}
    \includegraphics[width=\textwidth]{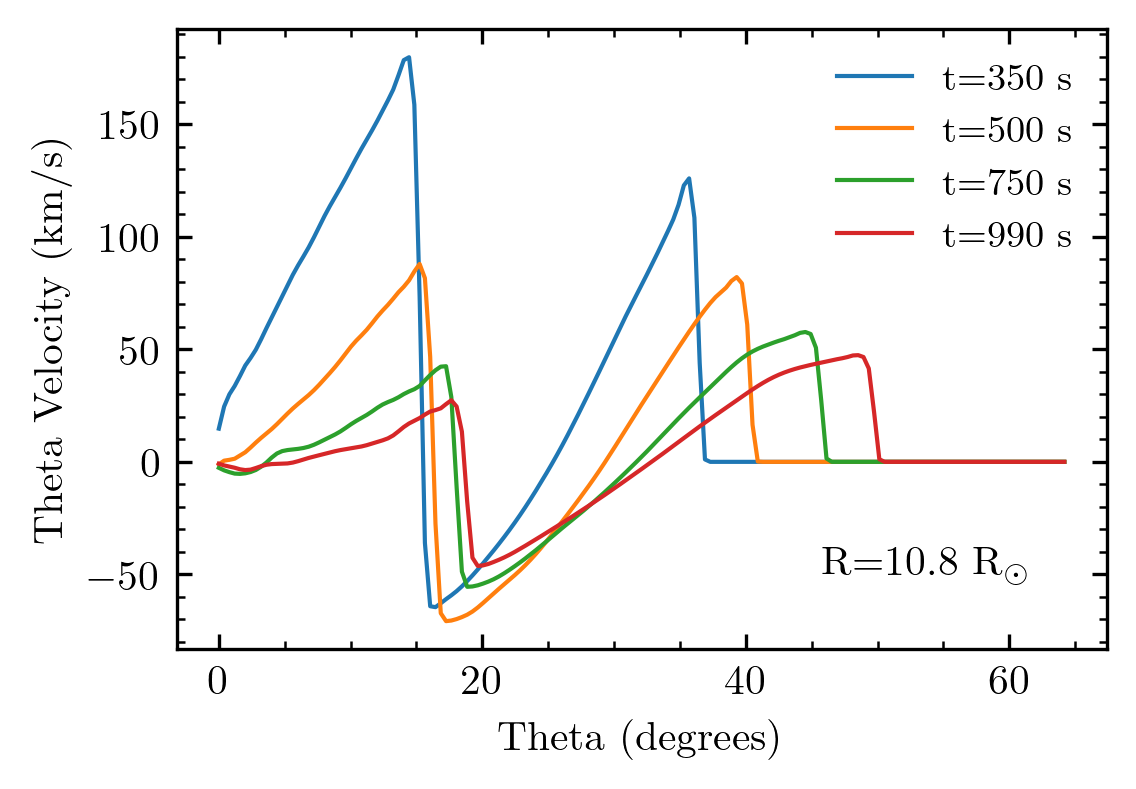}
  \end{minipage}

  \vspace{\floatsep} 

  \begin{minipage}[b]{0.48\textwidth}
    \includegraphics[width=\textwidth]{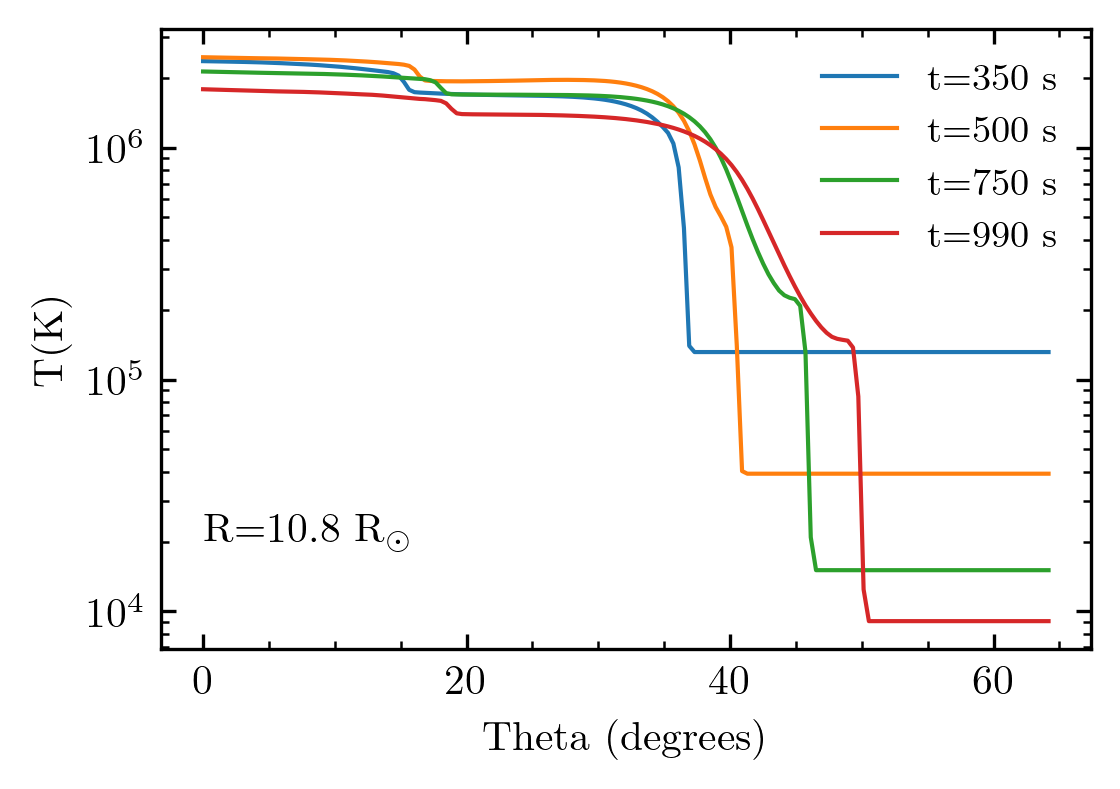}
  \end{minipage}
  \begin{minipage}[b]{0.48\textwidth}
    \includegraphics[width=\textwidth]{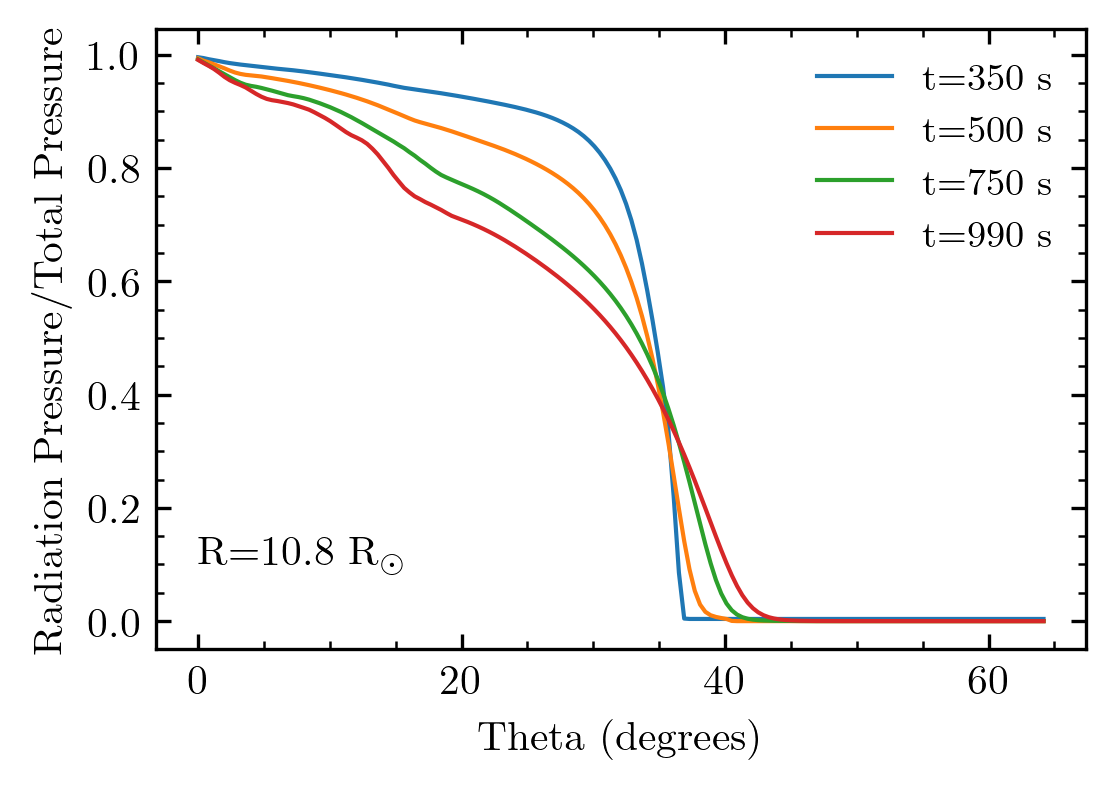}
  \end{minipage}
  \hfill

  \caption{Several properties of the ejecta measured at $r_{\rm out}=10.8~\rsun$ plotted vs $\theta$ at various times. The density (upper left) shows a rapid dropoff at small $\theta$, differing from \citet{2010ApJ...708.1025K}. The transverse velocity (upper right) illustrates the positions of the shocks and shows rarefaction waves between them. The temperature (lower left) reveals that the shock-heated ejecta is much hotter than the unshocked gas, causing much of the shocked ejecta to be radiation pressure-dominated (lower right). \label{fig:edgePlots2}}
\end{figure*}



We simulate the collision between the ejecta and the donor using the setup discussed above, terminating at $t=1000$ s. Slice plots of our results at $t=200$ s are shown in Fig.~\ref{fig:slicePlots}, wherein we see a bow shock adorning the donor. A weaker recompression shock also forms as ejecta converges on the $\theta=0$ pole downstream of the donor. These features lead to the formation of a low-density, high-temperature wake in the ejecta.

In Fig.~\ref{fig:edgePlots2} we show several properties of the ejecta as it outflows from the domain at $r_{\rm out}$. The density (upper left panel) shows discontinuities at both shocks, with the density dropping dramatically near the pole. This differs from \citet{2010ApJ...708.1025K}, where the recompression shock is absent and the density is uniform for small $\theta$. The low density within the wake affects the light curve, as we show in section \ref{sec:radiationResults}. The velocity in the $\theta$ direction reveals the presence of rarefaction waves between the shocks (upper right). We see that the temperature (lower left) is at least an order of magnitude greater for the shocked ejecta. The widening of the shock cone is also apparent in this plot, encompassing $\theta\lesssim 50^{\circ}$ by the end of the run. This means that roughly 18\% of the ejecta (by solid angle) has experienced shock heating. Furthermore, whereas the unshocked ejecta remains gas pressure-dominated, the shocked ejecta is primarily radiation pressure-dominated (lower right).

The use of a rigid boundary to model the donor allows us to directly compute the momentum transferred to the donor in the collision by simply integrating the pressure over the donor surface. Here the pressure is computed \change{within the Riemann solver} and is taken to be the momentum flux $\rho\vel\vel^{T}+P\mathbb{I}$ normal to the donor surface. We measure the total momentum imparted to the donor as $\Delta p=96$ $\msun$km/s, which exceeds the idealized estimate by $\approx$20\% (see Appendix \ref{sec:efficiency}). Though our donor is fixed to the grid, the kick velocity associated with this impulse would be $\Delta p / M_{d} = 279$ km/s, which is nearly perpendicular to the orbital speed $v_{\rm orb}=691$ km/s. This exceeds the kick velocities found in \citet{2019ApJ...887...68B} and \citet{Wong_2024} for comparable progenitors, both of which allowed for mass loss from the donor. The unbound mass carried away a portion of the imparted momentum which may otherwise have contributed to the kick velocity, suggesting that the treatment of the donor as a fixed, rigid sphere gives an upper limit on the imparted momentum.

\section{Evolving Optically Thick Ejecta} \label{sec:extrapolation}

Because the vast majority of the ejecta is optically thick for the first few hours following the supernova, it is undesirable to conduct a full 3-D radiation hydrodynamical simulation for this period. Instead, we evolve the ejecta forward in time assuming homology until the outer layers begin to radiate. Here we consider only the material which exited the outer boundary of the simulation discussed in the previous section, as the gas remaining inside the domain remains far inside the photosphere for the first day. 

We first confirm that the gas which passed through $r_{\rm out}$ is in homology by checking the ratio of gas plus radiation pressure to ram pressure, finding that $P/(\rho v^2/2)\ll 1$. We also check that the flow can be approximated as radial: the upper right panel of Fig.~\ref{fig:edgePlots2} shows that the transverse velocity is $\approx 100$ km/s, which is two orders of magnitude less than the radial velocity. Thus, the density of each Lagrangian parcel subsequently evolves as $\rho\propto t^{-3}$. The possible presence of $^{56}$Ni in the ejecta requires including the effects of radioactive heating, at a rate
\be
\epsilon=\frac{X_{56} E_{\rm nuc}}{A m_p \tau_{56}}e^{-t/\tau_{56}}, \label{radioactivity}
\ee
where \change{$X_{56}$ is the} mass fraction of $^{56}$Ni placed in our ejecta, $E_{\rm nuc}=1.72$ MeV is the energy released in the $^{56}\mathrm{Ni} \rightarrow ^{56}$Co decay, $\tau_{56}=757{,}728$ s is the e-folding time of $^{56}\mathrm{Ni}$ decay, and $A=56$ is its mass number. The temperature of each parcel then evolves as 
\be 
T\frac{ds}{dt}=\frac{X_{56} E_{\rm{nuc}}}{A m_p \tau_{56}} e^{-t/\tau_{56}}, \label{entropyChange}
\ee
where $s$ is the specific entropy (including radiation):
\be
s=\frac{k_B}{\mu m_p}\ln \lr{\frac{T^{3/2}}{\rho}}+\frac{4a_rT^3}{3\rho}. \label{entropyDensity}
\ee 
Equations \eqref{entropyChange} and \eqref{entropyDensity} are then solved for $T(t)$.

Our numerical scheme for evolving the fluid state of each parcel is as follows. We start at time $t_0=1000$ s with $s=s_0$, $T=T_0$, and $\rho=\rho_0$ and evolve to $t=7200$ s. We divide this time interval into $10^{4}$ equal time steps $\Delta t$, which is sufficient for convergence. For each time step, we evolve the parcel adiabatically and then update its entropy:
\begin{enumerate}
    \item Evolve the density as $\rho_{i+1}=\rho_0 \lr{t_0/t_{i+1}}^3$.
    
    \item Adiabatically evolve the gas to $t_{i+1}$ from $t_i$ by root finding (via the bisection method) for $T_{i+1}$ in \eqref{entropyDensity} with $\rho=\rho_{i+1}$ and $s=s_{i}$.
    
    \item Update the entropy according to \eqref{entropyChange}:
    \be
    s_{i+1}=s_i+\Delta t\lr{\frac{X E_{\rm nuc}}{A m_p \tau_{56}}\frac{2}{T_{i+1}+T_i}}e^{-t_i/\tau_{56}}.
    \ee
\end{enumerate}

\begin{figure}[t!]
  \centering
  \begin{minipage}[b]{0.48\textwidth}
    \includegraphics[width=\textwidth]{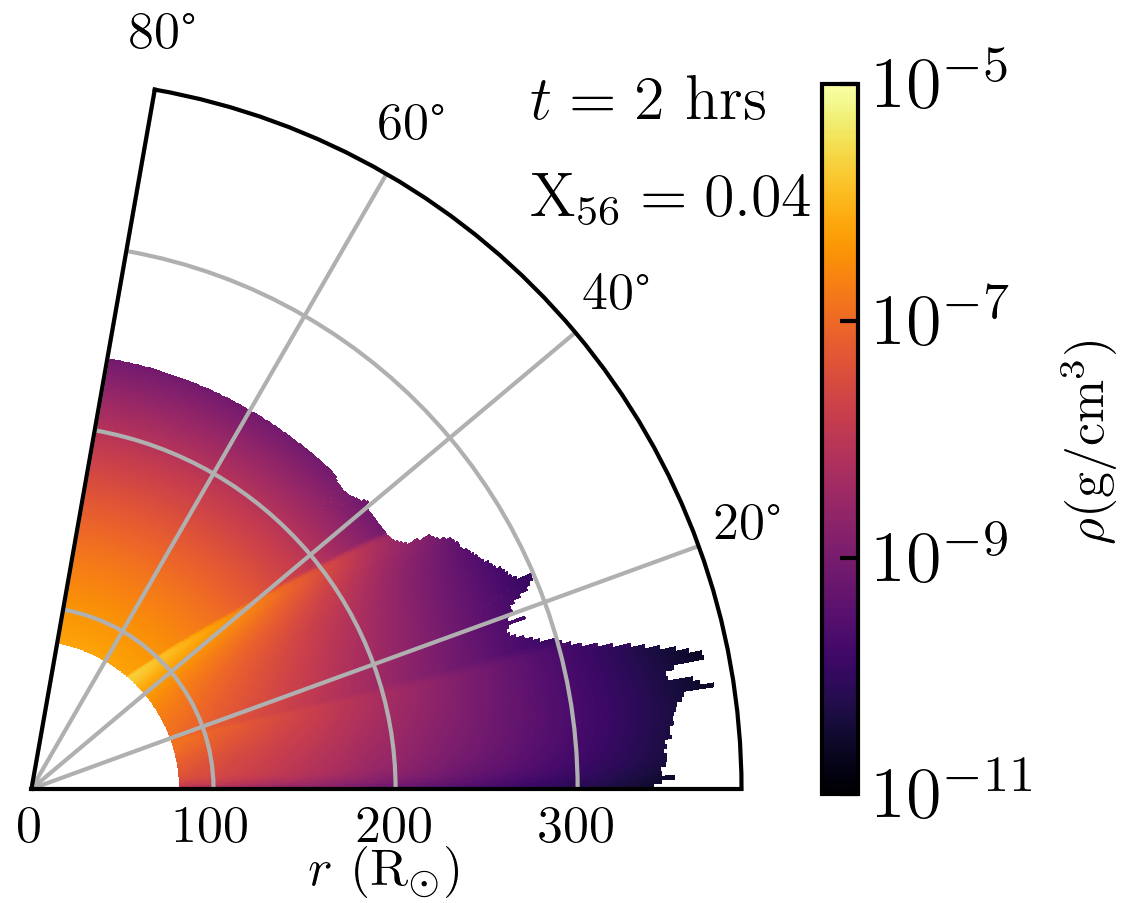}
  \end{minipage}
  \hfill
  \begin{minipage}[b]{0.48\textwidth}
    \includegraphics[width=\textwidth]{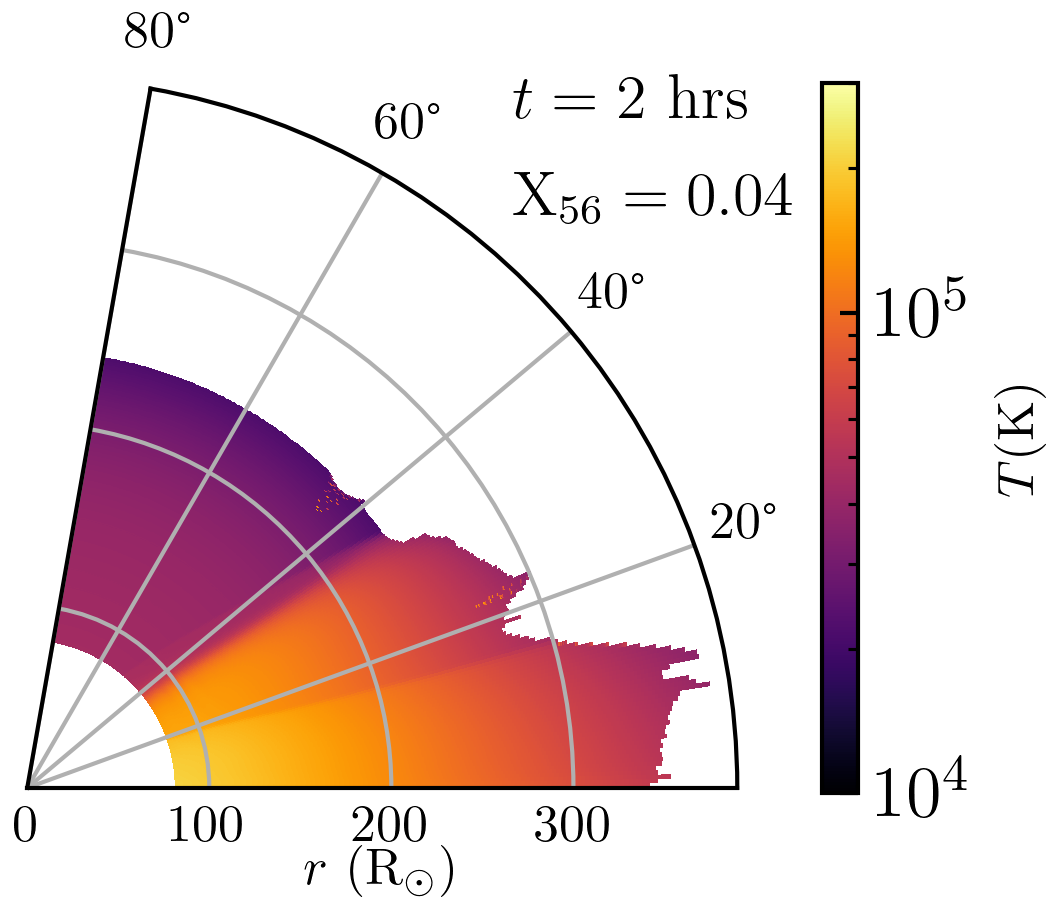}
  \end{minipage}

  \caption{Slice plots showing density (top) and temperature (bottom) on the $\phi=0$ plane after evolving the material leaving the edge of the original simulation to $t=2$ hours after the explosion for $X_{56}=0.04$.}
  \label{fig:evolvedStuff}
\end{figure}

\mycomment{
\begin{figure*}
  \centering
  \begin{minipage}[b]{0.48\textwidth}
    \includegraphics[width=\textwidth]{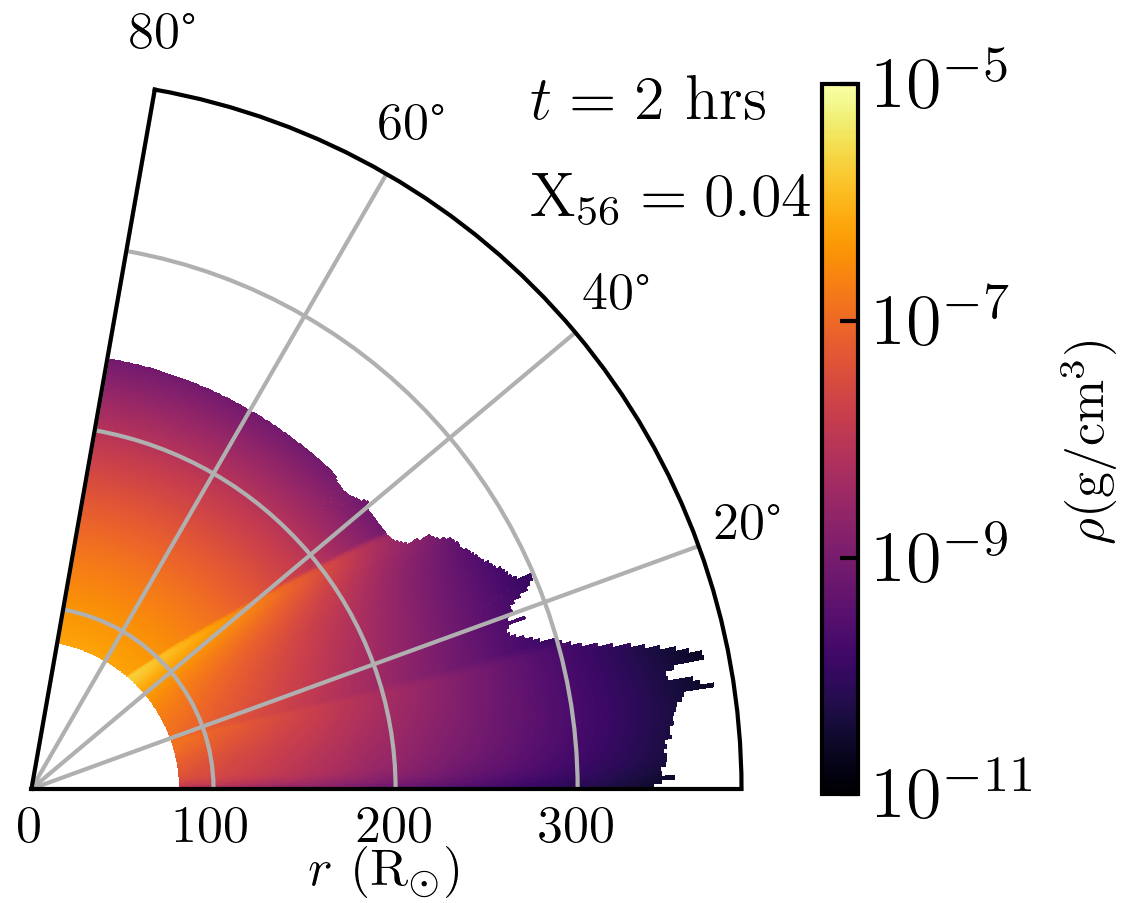}
    \label{fig:figure1}
  \end{minipage}
  \hfill
  \begin{minipage}[b]{0.48\textwidth}
    \includegraphics[width=\textwidth]{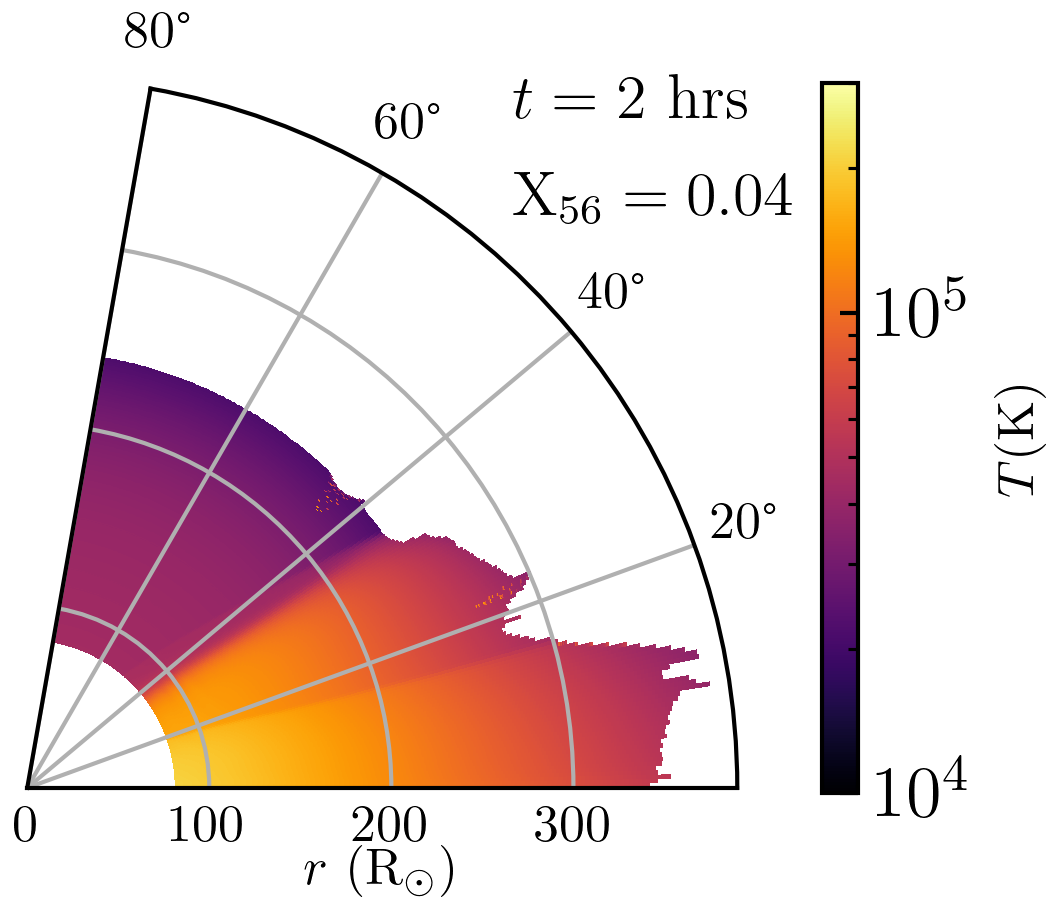}
    \label{fig:figure2}
  \end{minipage}

  \caption{Slice plots showing density (left) and temperature (right) on the $\phi=0$ plane after evolving the material leaving the edge of the original simulation to $t=7200$s after the supernova. }
  \label{fig:evolvedStuf}
\end{figure*}
}

This evolution yields a distribution of Lagrangian parcels, which we map back onto a spherical-polar grid to perform the radiation hydrodynamical simulation.  Here each grid cell adopted the fluid state of the Lagrangian parcel nearest to its cell center. The results of this effort are shown in Fig.~\ref{fig:evolvedStuff}, showing a clear distinction between unshocked ejecta, ejecta which was shocked by the bow shock, and ejecta which was shocked twice. Notably, the shocked ejecta protrudes far beyond unshocked ejecta due to the supersonic expansion on the downstream side of the donor, which accelerated the ejecta by as much as $\approx$50\% relative to the unshocked ejecta. As we will see in the following section, this means that the shocked ejecta is visible from a broad range of viewing angles. \change{We chose a nickel mass fraction of $X_{56}=0.04$ in the outermost ejecta, where different detonation models give more than an order of magnitude range for this value. As we are focused on early light curves the average nickel mass fraction is not of concern. When we refer to high-velocity ejecta we mean the ejecta up to $\theta\approx 20^{\circ}$ that protrudes in front of the unshocked material as shown in Fig.~\ref{fig:evolvedStuff}.} 

\section{Radiation and Light Curve} \label{sec:radiationResults}

To analyze the radiation emitted by the ejecta we perform radiation hydrodynamics simulations using Athena++ \citep{Jiang2021}, which solves the fluid equations coupled to radiation source terms. The time-dependent radiative transfer equation is solved iteratively, ensuring an accurate solution in both the diffusive and free-streaming limits. Additionally, the implicit solver described in \citet{Jiang2021} removes the speed of light as a timestep constraint, greatly reducing the computational cost. \change{The fluid evolves according to the Euler equations}
\be
\frac{\partial I}{\partial t}+c \bold{n}\cdot \bold{\nabla I}&=&c(\eta-\chi I), \label{radTransfer} \\
\frac{\partial \rho \bold{v}}{\partial t}+\nabla \cdot \rho \bold{v}\bold{v}+\nabla P&=&-\bold{S_r(P)}, \label{momentumWithRad} \\
\frac{\partial u_g}{\partial t}+\nabla \cdot (u_g+P)\bold{v}&=&-S_r(u_g),\label{energyWithRad}
\ee
\change{and is coupled to the radiation field via the energy and momentum source terms}
\be
S_r(E)&=&4\pi c \int S_I d\Omega, \\
\bold{S_r(P)}&=&4\pi \int \bold{n}S_I d\Omega.
\ee
\change{These are moments of the frequency-integrated source term}
\begin{dmath}
S_I=\rho\Lambda^{-3}\lrb{(\kappa_s+\kappa_a)(J_0-I_0)+(\kappa_a+\kappa_{\delta P})\lr{\frac{a_rT^4}{4\pi}-J_0}},
\end{dmath}
\change{where}
\be
\Lambda(\bold{n}, \bold{v})&=&\gamma(1-\bold{n}\cdot \bold{v}/c).
\ee
Here $c$ is the speed of light, $\eta$ is the lab frame emissivity, $\chi$ is the lab frame opacity, $u_g$ is the internal energy of the gas, and $I$ is the frequency-integrated specific intensity of the radiation field which is a function of space, time, and angular direction $\bold{n}$. In the source terms $\gamma$ is the Lorentz factor, $\kappa_s$ is the scattering opacity, $\kappa_a$ is the Rosseland mean absorption opacity, and $\kappa_{\delta P}$ is the difference between the Planck mean and Rosseland mean opacities. The variables $I_0$ and $J_0$ are the radiation specific intensity and flux in the comoving frame of the ejecta, respectively. For the opacity, we want $\kappa_s+\kappa_a$ (absorbing + scattering) to be equal to $0.2$ $\rm{cm}^2/\rm{g}$ for Thompson scattering in the absence of hydrogen. We choose $\kappa_s=0.17$ $\rm{cm}^2/\rm{g}$ and $\kappa_a=0.03$ $\rm{cm}^2/\rm{g}$. The artificially high absorption scattering ensures that gas and radiation are strongly coupled in our simulation while ensuring that the total opacity is $0.2$ $\rm{cm}^2/\rm{g}$. We implement radioactive heating from the $\isotope[56]{Ni}$ decay chain by adding a local energy source term.

As kinetic energy accounts for the vast majority of the explosion energy, numerical errors during inversions between the gas pressure and the gas total energy can cause erroneous conversions of kinetic to thermal energy. This excess heating then leaks into the radiation field and causes errors. To combat this, we use a high spatial resolution on a small domain spanning $r=172.6~\rsun$ to $r=1{,}649.8~\rsun$. At $t=8$ hrs we then re-map the ejecta and the radiation field onto a larger grid spanning $r=710.7~\rsun$ to $r=4{,}822.4~\rsun$ allowing the integration to continue to a final time of $t_f=23.9$ hrs. For both of these simulations a radial scaling factor of $x=1.0012$ was used to construct the mesh.

We initialize our radiation hydrodynamics simulations by depositing the Lagrangian fluid elements described in the previous section onto a spherical-polar grid via a nearest neighbor search. The simulation now spans the full range of $\theta$ and $\phi$ with the supernova at the origin so as to generate synthetic light curves for observers at all angles.
The radiation field is additionally specified using 80 angles at every point whose discretization scheme is described in \citet{Jiang2021}. For this run the domain contains $n_r\times n_{\theta} \times n_{\phi}=2304\times 300 \times 30=20{,}736{,}000$ cells.

We set the ambient medium to have density $\rho=2.42\times 10^{-22}$ $\rm{g}/\rm{cm}^3$ and pressure $P=4.13\times 10^{-17}$ $\rm{ergs}/\rm{cm}^3$. The low pressure ensures that the ambient medium generates a negligible amount of radiation. When the ejecta collides with the ambient medium it creates an energy flux of $\rho v^3/2$ which leads to a luminosity of $\sim \rho v^3 r^2$, which is negligible given our choice of density. 

Boundary conditions were set to allow radiation and fluid to freely exit through the outer radial boundary. At the inner radial boundary we insert material with the same properties ($\rho$, $v_r$, etc...) as the material already at the inner boundary. This ensures that the fluid that was initialized into the simulation does not experience abnormal pressure gradients and flows outwards as expected. This material remains optically thick and does not affect the light curve during the first day. 

\subsection{Simulation Results}

\begin{figure}
  \centering
\includegraphics[width=\linewidth]{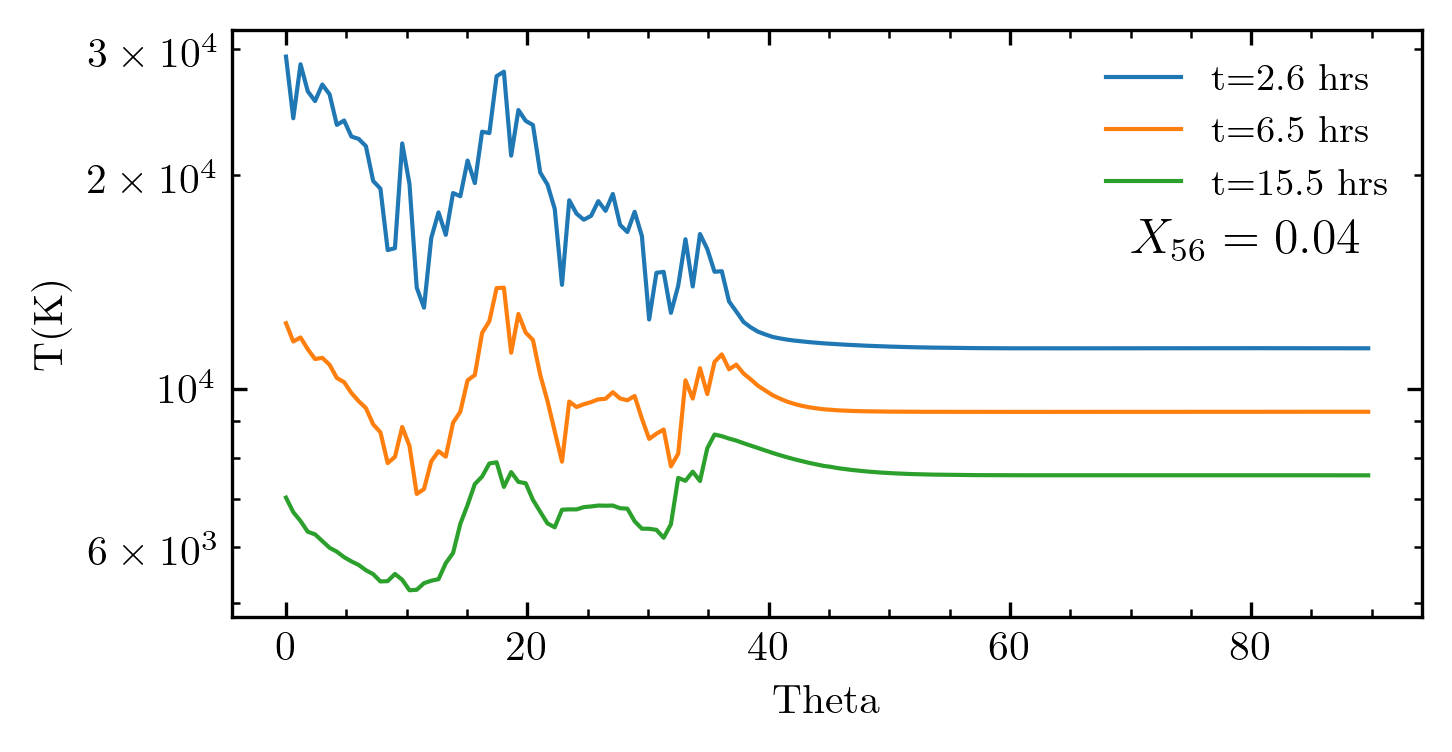}
\caption{Temperature as a function of $\theta$ at $\tau=1$ (integrated radially) for a set of times after the explosion.}
\label{fig:photosphere}
\end{figure}

\begin{figure}[t!]
  \centering
  \begin{minipage}[b]{0.48\textwidth}
    \includegraphics[width=\textwidth]{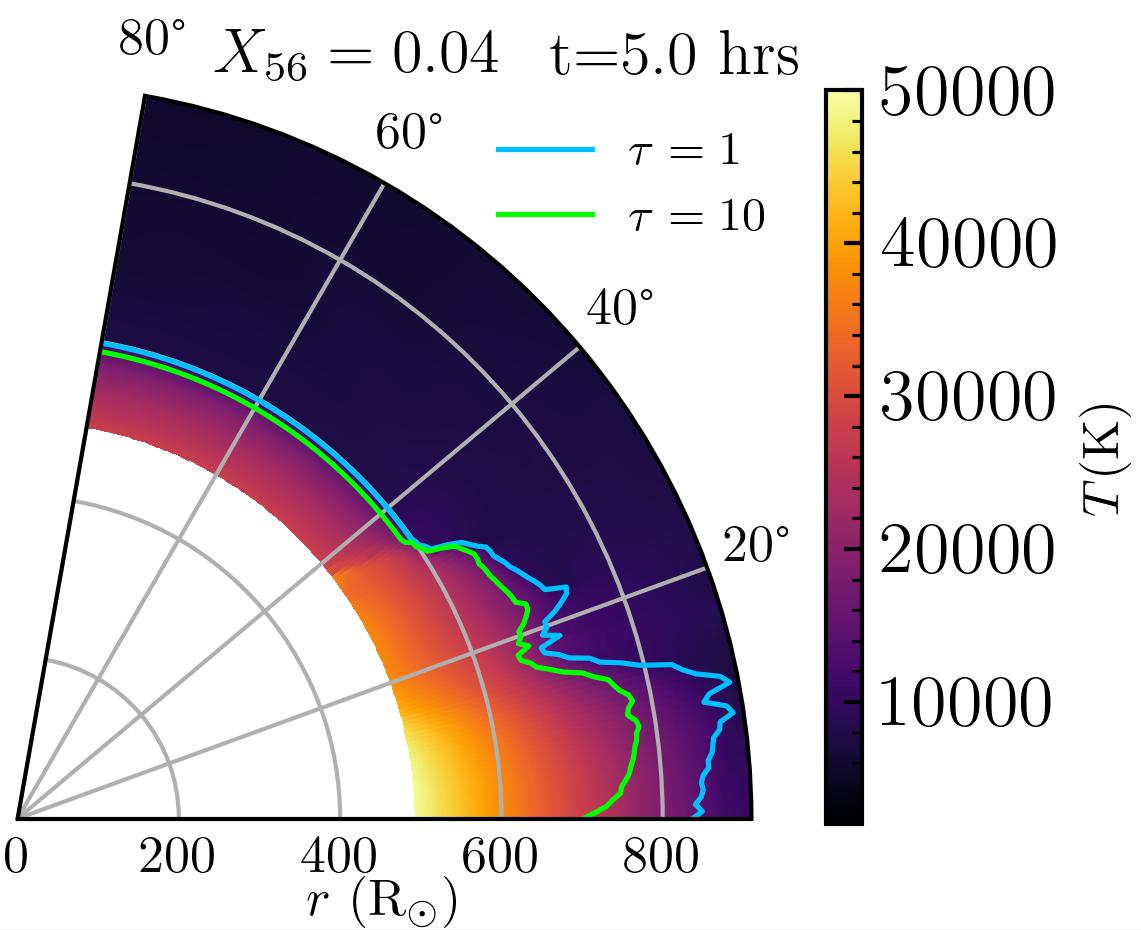}
    \label{fig:figure1}
  \end{minipage}
  \hfill
  \begin{minipage}[b]{0.48\textwidth}
    \includegraphics[width=\textwidth]{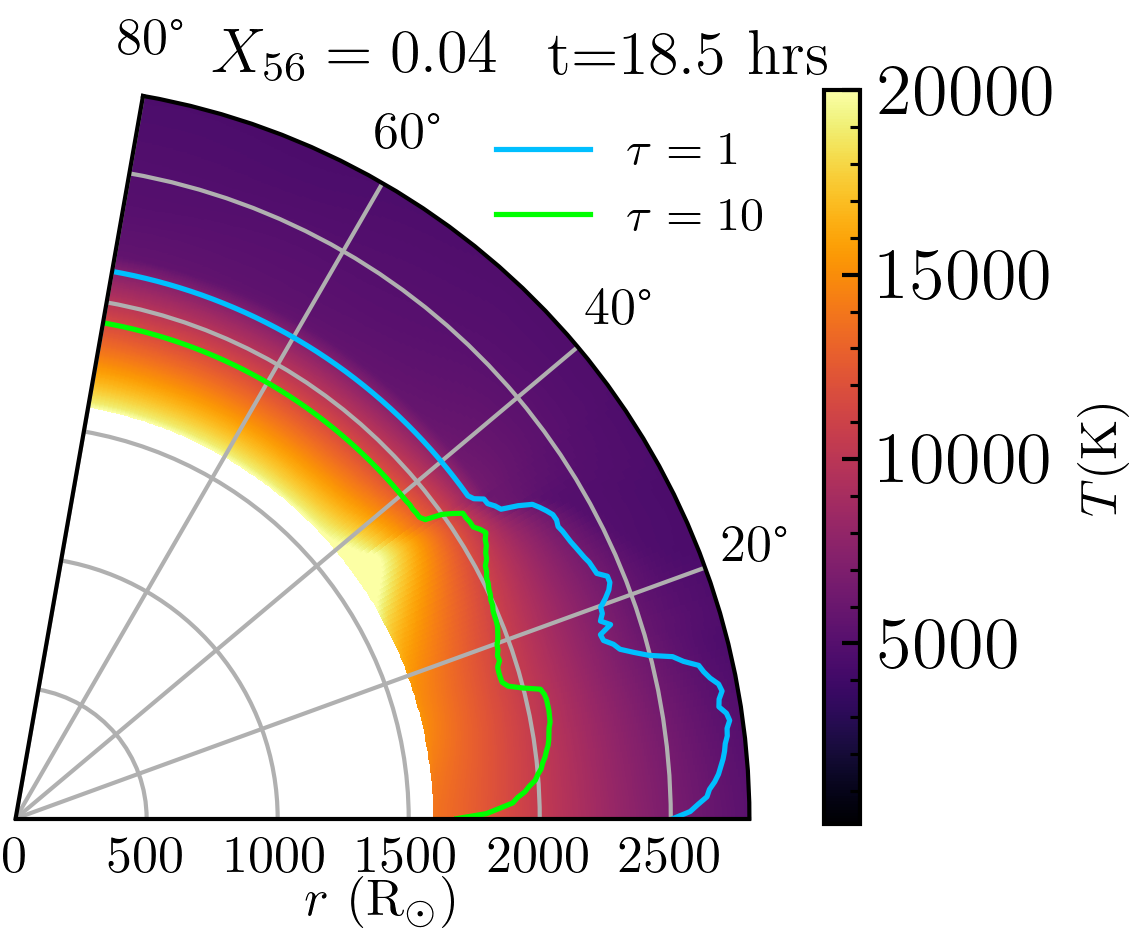}
    \label{fig:figure2}
  \end{minipage}

  \caption{Plots showing temperature and the $\tau=1$ and $\tau=10$ optical thickness isolines on the $\phi=0$ plane at $t=5.0$ hrs (top) and $t=18.5$ hrs (bottom) assuming $X_{56}=0.04$.}
  \label{fig:tauIsolines}
\end{figure}

We first consider the photosphere and how it evolves with time. Fig.~\ref{fig:photosphere} shows the temperature of the photosphere at $t=2.6$, $6.5$, and $15.5$ hrs. The wake drops in temperature more quickly than the unshocked material due its lower density. The unshocked material is initially gas-pressure dominated so that the radioactive heating has a larger effect on its temperature than in the shocked material, which is radiation-pressure dominated (both due to its higher temperature and lower density). The full temperature profiles at two representative times of $t=5.0$ and $18.5$ hrs are plotted in Fig.~\ref{fig:tauIsolines} as well as isolines of radially integrated optical depth $\tau$ to illustrate the geometry and temperature of the photosphere and how it changes with time. The photosphere becomes more uniform in radius as time progresses. 

\change{Though we do not calculate line profiles in this work, it is evident in the bottom panel of Fig.~\ref{fig:tauIsolines} that observers at angles exceeding that of the wake will measure the photosphere within the wake at higher velocities than the unshocked ejecta. This effect is of course most pronounced for observers at small $\theta$ where the projected velocity is the highest. In the absence of line profile calculations we choose not to speculate about the observability of these features.}

\begin{figure}
    \includegraphics[width=\linewidth]{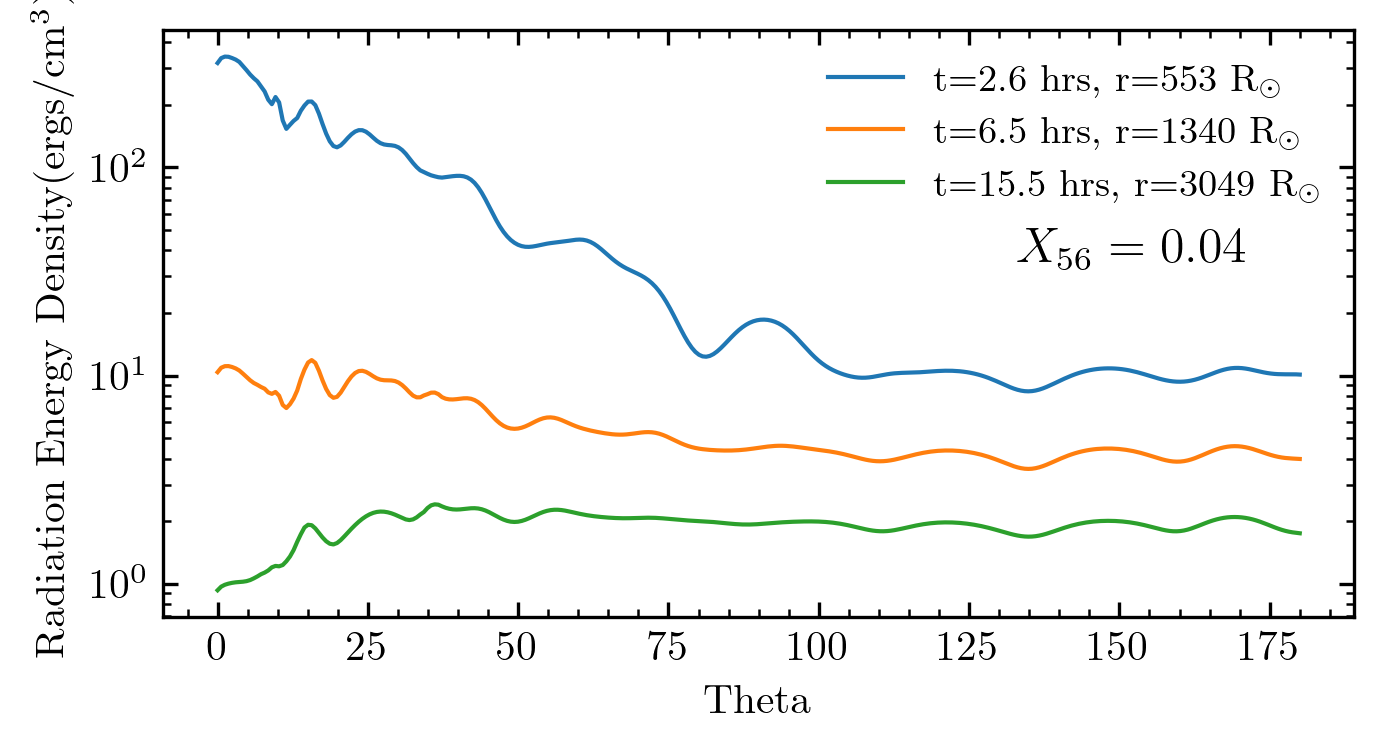}
 
  \caption{Radiation energy density at the measurement radius $R_{m}$ vs $\theta$ at various times. }
   \label{fig:radiationEnergyDensity}
\end{figure}

To generate synthetic light curves, the intensities along all 80 angles are stored at every timestep at a measurement radius $R_m(t)=v_0t+R_0$, where $R_0=431.5~\rsun$, $v_0=37{,}000$ km/s, and $t$ is measured with respect to the start of the simulation. This places $R_m(t)$ safely outside of the ejecta. The radiation energy density at this radius which is plotted in Fig.~\ref{fig:radiationEnergyDensity}, from which it is evident that the shocked material produces more radiation than the unshocked material with the amount steadily decreasing with $\theta$. The noise in Fig.~\ref{fig:radiationEnergyDensity} is due to the angular discretization scheme used in Athena++.
Most importantly, the radiation from the shocked material extends to angles much larger than that of the wake itself due to the radial protrusion of the shocked ejecta, as is visible in Fig.~\ref{fig:evolvedStuff}.

\subsection{Light Curves for all Observers}

We now determine the luminosity seen by observers at various angles. For an observer viewing from a direction $\hat{n}_O$, the luminosity is given by
\begin{equation}
    L=4\pi R^2 \int_S I(R,\theta, \phi,\hat{n}_O) (\hat{r} \cdot \hat{n}_O)\chi(\hat{r}\cdot \hat{n}_O) d\Omega,
    \label{luminosityMain}
\end{equation}
where $I(R,\theta,\phi, \hat{n}_O)$ is the intensity of the radiation field at the point $(R, \theta, \phi)$ radiating into the direction $\hat{n}_O$, $\chi$ is the step function, and $S$ is the sphere where intensities were recorded.

To ensure that the simulation is self-consistent, we also check that the radiation flux $F$ at the photosphere results in a luminosity $L=4\pi R^2 F$ that is consistent with Eq.~\ref{luminosityMain}. As the photosphere and measurement radius are only a few light-seconds apart, this comparison is legitimate. The two luminosities are equal to within a few percent. As a second check, we compare the comoving flux output by Athena++ to the slope of the radiation energy density, $(d(caT^4)/d\tau)/3$, for material with $\tau>10$. We find excellent agreement, confirming that Athena++ properly captures behavior in the diffusive limit.

\begin{figure*}
  \centering
\includegraphics[width=\linewidth]{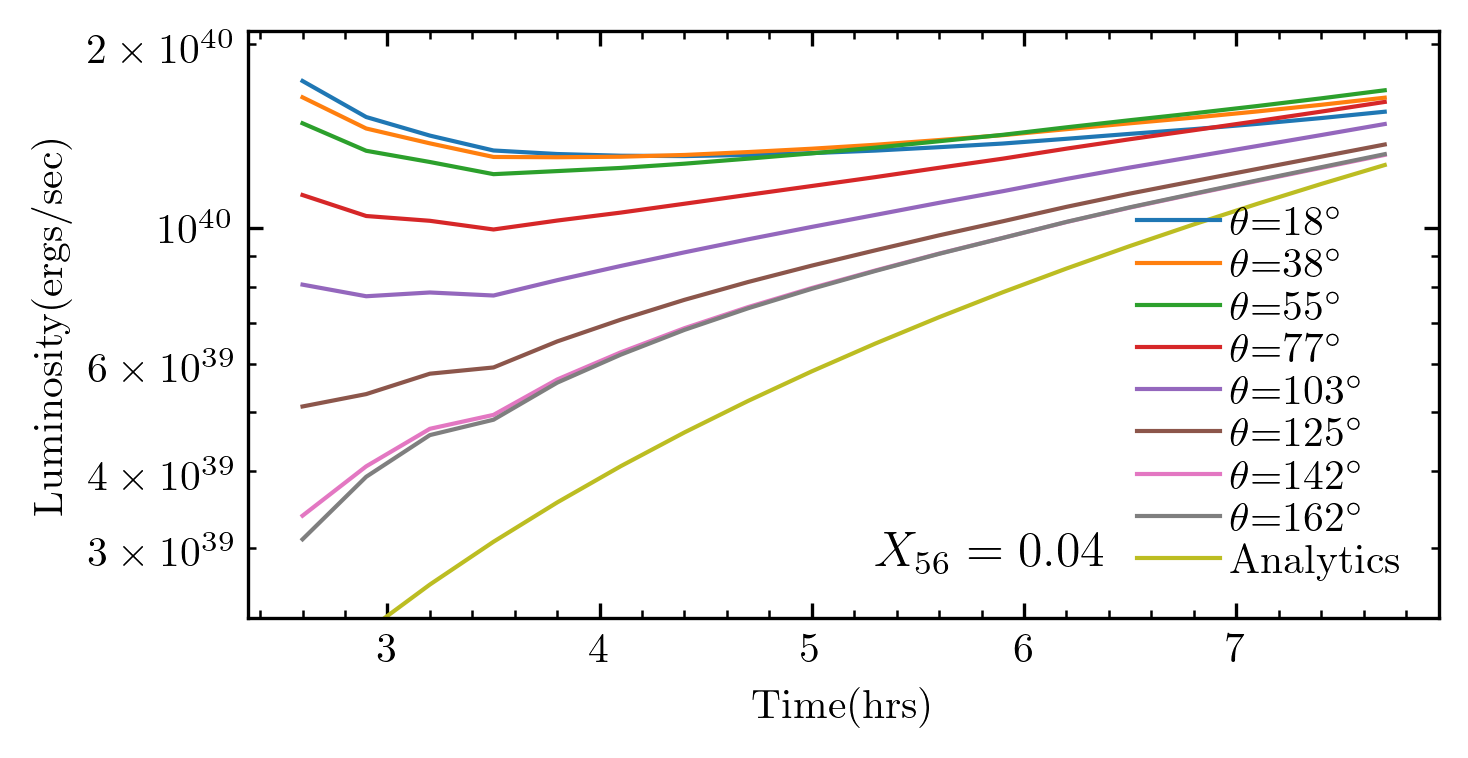}
\caption{Luminosity vs time for $X_{56}=0.04$ from $t\approx2$ to 8 hrs. The legend provides the angle from which the observer views the supernova. We compare against the light curve predicted by \citet{2012ApJ...759...83P} labeled ``Analytics'' describing a supernova without shock-heated ejecta.}
\label{fig:firstSixCurve}
\end{figure*}

\begin{figure*}
  \centering
\includegraphics[width=\linewidth]{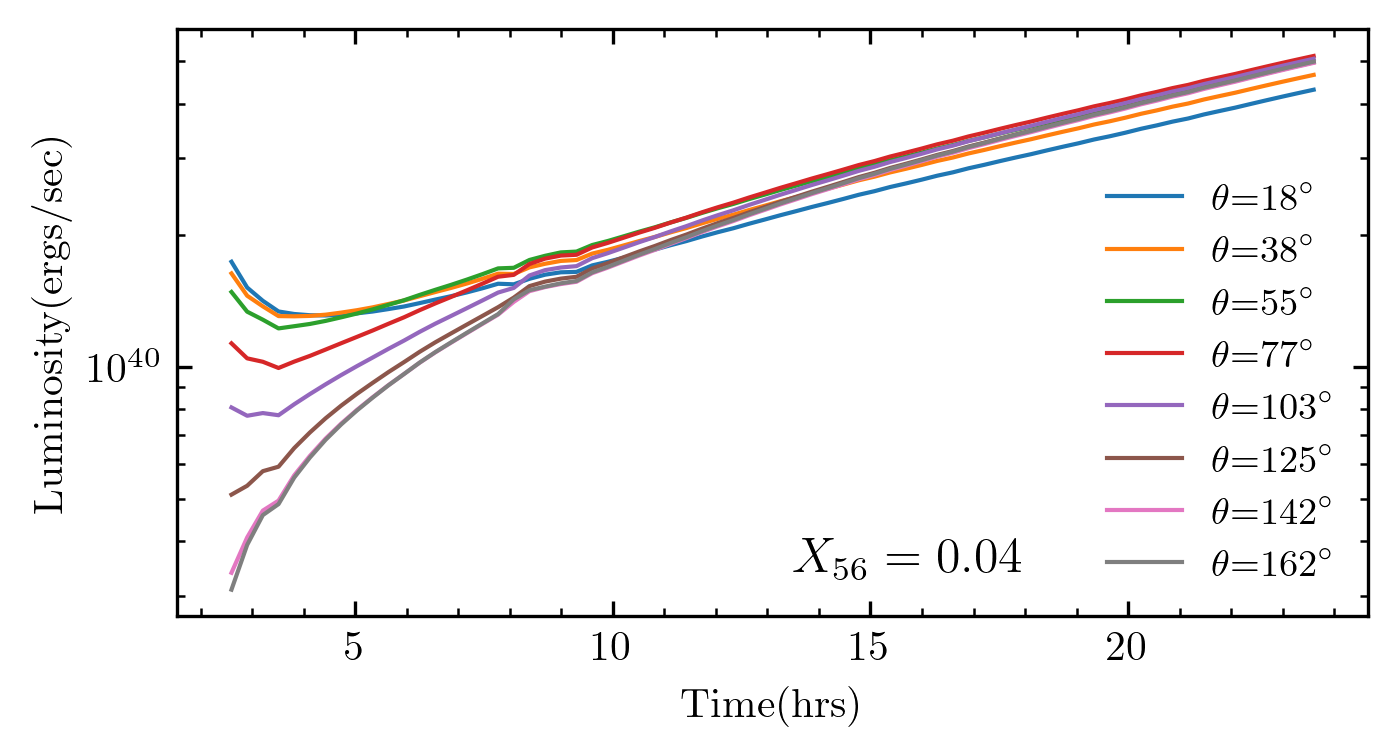}
\caption{Luminosity vs time for $X_{56}=0.04$ up to $\approx$1 day after the supernova at various viewing angles. }
\label{fig:fullDayCurve}
\end{figure*}

\begin{figure*}
  \centering
\includegraphics[width=\linewidth]{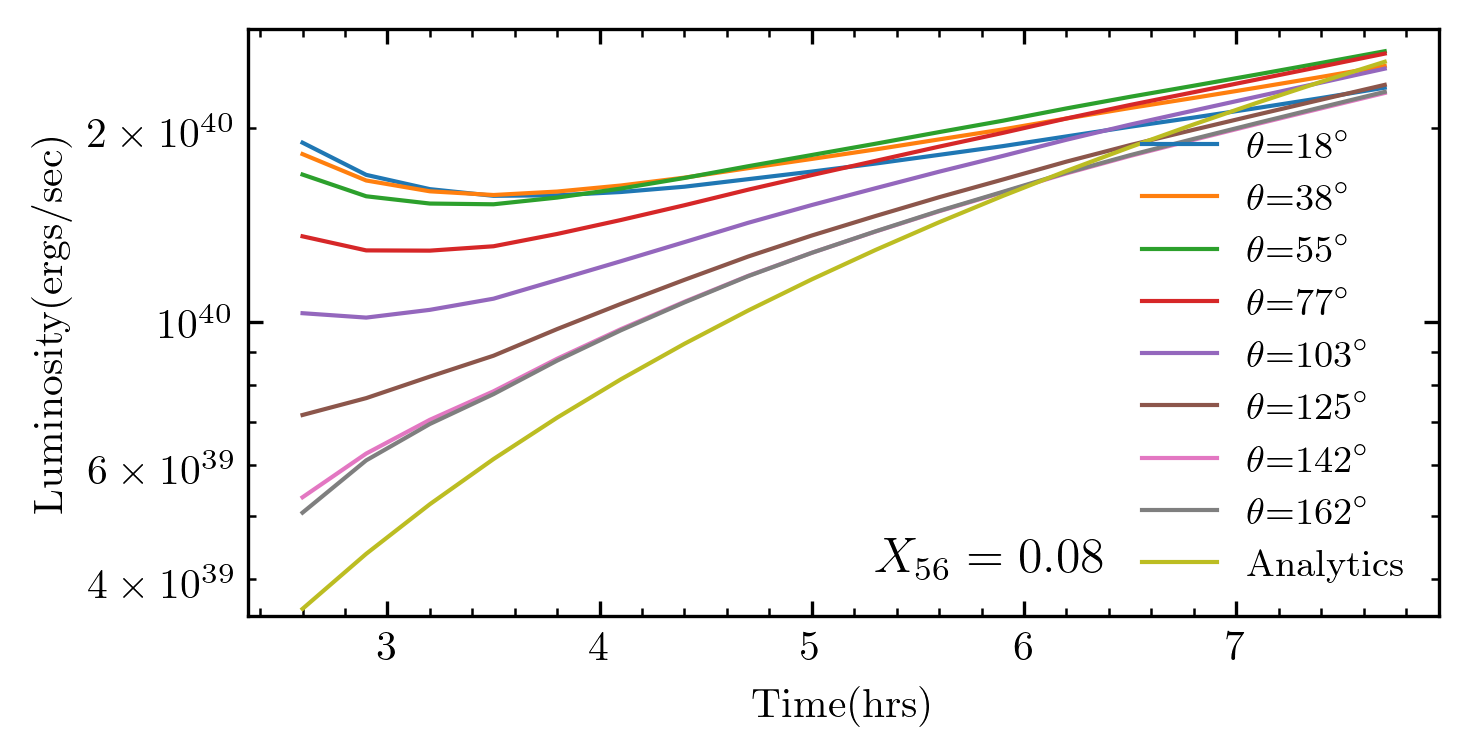}
\caption{Luminosity vs time for $X_{56}=0.08$ from $t\approx2$ to 8 hrs. The legend provides the angle from which the observer views the supernova. We compare against the light curve predicted by \citet{2012ApJ...759...83P} labeled ``Analytics'' describing a supernova without shock-heated ejecta.}
\label{fig:firstSixEight}
\end{figure*}

Each of the angular ordinates for the radiation field is specified by a $(\theta_k, \phi_k)$. We sort these ordinates into 8 distinct bins in $\theta$, where inside each bin the $\theta_k$ values deviate from one another by no more than $1.8^{\circ}$ with each bin having a mean $\theta$ value $\theta_i$. We calculate $L(t)$ for an observer at each of the 80 ordinates by setting $\hat{n}_O$ to the unit vector specified by $(\theta_k,\phi_k)$. Then, for all $\theta$ bins, we average the luminosity of the observers within them to obtain $L(t)$ for an observer at $\theta_i$. 

We show the full family of light curves from 2 to 8 hours after the explosion in Fig.~\ref{fig:firstSixCurve}, as well as an analytical model from \citet{2012ApJ...759...83P}. Observers at large $\theta$ are seeing the 
``back side" of the explosion, opposite to that of the wake. In the absence of a companion interaction, the early light-curve of Type Ia SNe depends 
on the abundances and locations of radioactive elements in the outermost ejecta \citep{2021MNRAS.502.3003R, 2020A&A...642A.189M, 2020A&A...634A..37M, 2018A&A...614A.115M, 2017MNRAS.472.2787N,2016ApJ...826...96P,2014ApJ...784...85P, 2013ApJ...769...67P, 2012ApJ...759...83P}, as that heating maintains the high temperatures in the ejecta during homologous expansion. \citet{2012ApJ...759...83P} studied the case closest to ours---a uniform $^{56}$Ni distribution---making a prediction plotted with the label ``Analytics'' which is in reasonable agreement with our results given the differing assumptions in the ejecta density profile. For observers able to see the wake, the luminosity is initially high, specifically for an observer viewing from $\theta=18^{\circ}$ it is $1.52\times 10^{40}$ ergs/s at $t=2.9$ hrs as compared to the $3.92\times 10^{39}$ ergs/s that an observer viewing from the backside sees. For the observers that see the wake this luminosity decays with time, in agreement with \citet{2010ApJ...708.1025K}. Though \citet{2010ApJ...708.1025K} focused on larger orbital separations, his equation (22)  predicts a luminosity of $2.6 \times 10^{41}$ ergs/s at $t=2.9$ hrs, roughly an order of magnitude larger than what we find at this time. At early times, $L(t)$ decreases with $\theta$ as the photosphere recedes and the observer sees less of the shocked material.
However, the effect of the shock can still be seen at quite large angles up to $\theta=142^{\circ}$. This means that while the strongest features of the effect are visible to around $\theta=55^{\circ}$ (roughly 20\% of the sky), the effect can still be seen to a moderate degree over more than half of the sky. 


At later times, the angular contrast fades, and eventually all material approaches a rising light curve driven by radioactive heating. The light curve for $t=2$ hrs through $t=24$ hrs is plotted in Fig.~\ref{fig:fullDayCurve}. Here all viewing angles at sufficiently high $\theta$ ($\gtrsim 50^{\circ}$) eventually converge to nearly identical solutions. However, the observers looking directly into the wake see the luminosity dip below that of the unshocked material in an inversion of the behavior at earlier times. We find that observers at $\theta=18^{\circ}$ at $t=23$ hrs see $86\%$ of the value seen by observers viewing the back side of the collision, 
and observers at $\theta=38^{\circ}$ see $93\%$ of this value. This persists for the remainder of our integration, and for angles less than $\approx$$50^{\circ}$ the observed luminosity is persistently $\approx$15\% lower than from the back side. This is due to the modified density structure within the wake, which affects both the radioactive heating and the subsequent radiation transfer. Whether this persists up to the peak of the light curve is a question we leave to future work.

\change{We also studied a higher Nickel mass of $X_{56}=0.08$ for the purpose of comparison; the resulting light curves from 2 to 8 hours after the explosion are plotted in Fig.~\ref{fig:firstSixEight}. Many of the features of the $X_{56}=0.04$ case persist---in particular, the light curves for angles smaller than $\theta\approx 50^{\circ}$ are nearly identical between the two nickel masses, as the thermodynamics of this material is dominated by the shock. The clearest difference is that the nickel heating becomes dominant at earlier times for observers at small $\theta$. At $t>8$ hrs the wake appears dimmer than the unshocked material as in the $X_{56}=0.04$ case, indicating that this feature is most likely due to the density structure of the material.}

\section{Discussion \& Conclusions} \label{sec:discussion}

In this paper we explored the ejecta interaction with the donor star in a double degenerate type Ia supernova and its \change{effect} on the ejecta structure and early light curve. The shock heating leads to observers seeing the shocked material as brighter than the unshocked material for at least the first day. The ejecta morphology was also explored, finding that the nature of the shocks behind the donor leads to ejecta moving faster in the wake, getting ahead of the unshocked ejecta. This allows for observers at large angles to the wake to see the brighter emission from the shocked ejecta, reaching over one half of the sky for the first few hours. 

For the purposes of this initial exploration, we placed a uniform distribution of $^{56}$Ni within the ejecta, enabling us to determine when the shock-heating effects diminish relative to the intrinsic light curve expected from the explosion.  
After about 12 hours, we find that the nickel heating  dominates the ejecta, so that most observers see nearly the same luminosity, which is in good agreement with the analytical predictions of \citet{2012ApJ...759...83P}. However, the modified density profile in the wake leads to a 15\% underbrightness for observers directly in the wake compared to that seen in the unshocked ejecta. Whether this persists to 
later times near the peak of the light curve remains an open question.

Hence, there are two ways the effect of this collision can be seen. First, in the very early time light curves there will be an overbrightness which we calculate to be $1.5\times 10^{40}$ ergs/s at $t=2.9$ hrs for the observer viewing from $\theta=18^{\circ}$ as compared to the $3.9\times 10^{39}$ ergs/s that an observer viewing from the backside sees at that time for our assumed nickel mass. This overbrightness is roughly an order of magnitude smaller than that extrapolated from the work of \citet{2010ApJ...708.1025K}. This overbrightness decays away after about 12 hours. Second, at late times (i.e.~one day) the interaction still has an effect, as observers at $\theta=18^{\circ}$ then see $86\%$ of the value seen by observers viewing the backside of the collision of $4.9\times 10^{40}$ ergs/s, and observers at $\theta=38^{\circ}$ see $93\%$. This is due to the permanent marks on the ejecta density structure from the donor interaction. This may imply that the effects of the collision on creating the wake continue to impact the later light curves of Type Ia supernovae. 


Though UV observations of SNIa have largely ruled out red giant or red supergiant donors \citep{2012ApJ...749...18B}, they allow for the possibility of binaries with smaller orbital separations corresponding to main-sequence or degenerate donors \citep{2015PASJ...67...54K}. Alternatively, SNIa which show no sign of donor interaction may be merger products \citep{2015Natur.521..332O}. However, the high-velocity ejecta found in this work may provide a mechanism for the two distinct ejecta velocities found in SN 2021aefx, which also contained the fastest ejecta observed in an SNIa to date \citep{2023ApJ...959..132N}. SN 2021aefx is a member of a subclass of SNIa which exhibit excess emission in the blue band over the first 0.5 to 2 days \citep{2025ApJ...983....3N}. Furthermore, \citet{2025ApJ...984..160I} conclude that in SN 2021hpr, early-excess emission is associated with high-velocity features in the Si and Ca lines. \citet{2025arXiv250217556H} found similar features in SN 2024epr, noting that neither a delayed detonation nor a thin shell He detonation sufficiently reproduced its observed properties. 

\change{The $\approx 2\times 10^{40}$ ergs/s luminosity associated with the shocked ejecta in the double degenerate scenario is more than 100 times lower than that predicted by \citet{2010ApJ...708.1025K} for red giant companions. This creates a special challenge to detecting the signal, as very few SNe are detected at bolometric luminosities fainter than the $M_{\rm Bol}\approx -12$ implied in our work. To date, only SNe 2018aoz has been detected at these low brightnesses in multiple bands \citep{2022NatAs...6..568N,2023ApJ...946....7N}. That landmark work found brightnesses consistent with our calculations as a function of time. We cannot yet calculate colors with our grey radiative transfer approach and cannot speak to consistency with the observed color evolution.  In future work, multi-group 3D radiation hydrodynamics calculations will need to be carried out to fully test against the existing data and any future events.}

\begin{acknowledgments}
\change{We thank the anonymous reviewer for their constructive suggestions which improved the manuscript.} This research benefited from collaborations funded by the Gordon and Betty Moore Foundation through Grant GBMF5076. This work was performed in part at the Aspen Center for Physics, which is supported by National Science Foundation grant PHY-2210452. This research was supported in part by grant NSF PHY-2309135 to the Kavli Institute for Theoretical Physics (KITP). GK and LJP are supported by grant ATP-80NSSC22K0725 through the NASA Astrophysics Theory Program. LJP is also supported by a grant from the Simons Foundation (216179, LB). We use the Matplotlib \citep{Hunter:2007} and SciPy \citep{2020SciPy-NMeth} software packages for the generation of plots in this paper. Use was made of computational facilities purchased with funds from the National Science Foundation (CNS-1725797) and administered by the Center for Scientific Computing (CSC). The CSC is supported by the California NanoSystems Institute and the Materials Research Science and Engineering Center (MRSEC; NSF DMR 2308708) at UC Santa Barbara.
\end{acknowledgments}


\software{Athena++ \citep{athena++},
          Matplotlib \citep{Hunter:2007},
          NumPy \citep{harris2020arrayNUMPY},
          SciPy \citep{2020SciPy-NMeth}
          }

\newpage
\appendix
\section{Momentum Transfer Efficiency} \label{sec:efficiency}

\citet{2018ApJ...864..119H} computed the efficiency of momentum transfer from planar ejecta onto a spherical donor, assuming that the momentum $\mom$ of an ejecta particle was deflected tangent to the surface of the donor at the point of intersection. Under these idealized assumptions they calculated that the fraction of incoming momentum imparted to the donor is $\eta_{\rm ideal}=1/2$, though in practice they found that $\eta$ can vary from 0.17 to 0.72 depending on the hydrodynamic response of the donor. We extend this formalism to spherically-symmetric ejecta.

\begin{figure}
\centering
  \includegraphics[width=0.45\textwidth]{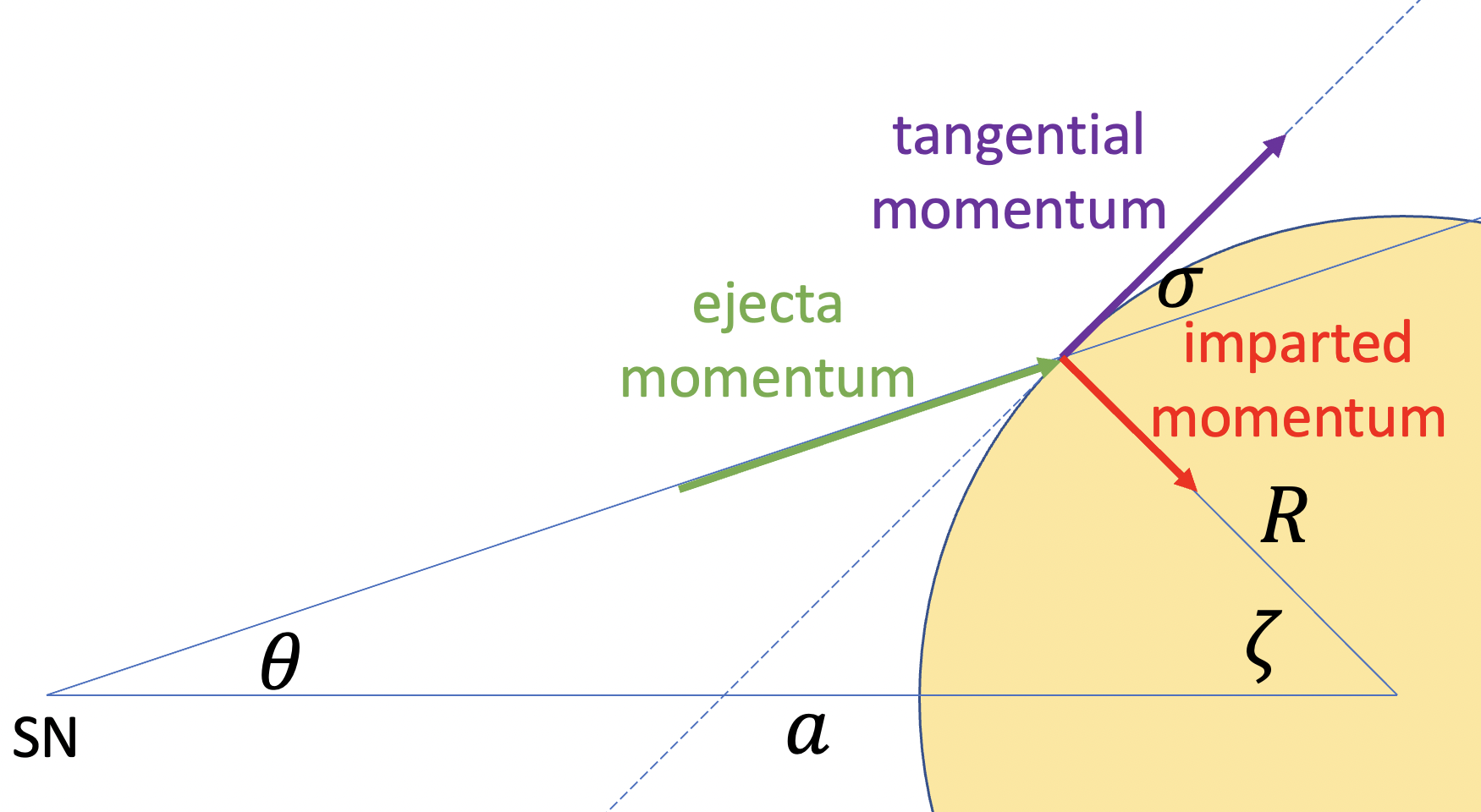}
    \caption{Illustration of spherically-symmetric ejecta impacting a spherical donor, with variable definitions. The ejecta momentum is decomposed into its components tangential and normal to the donor surface, the latter being imparted to the donor.
    \label{fig:momentumdiagram}}
\end{figure}

\begin{figure}
\centering
  \includegraphics[width=0.45\textwidth]{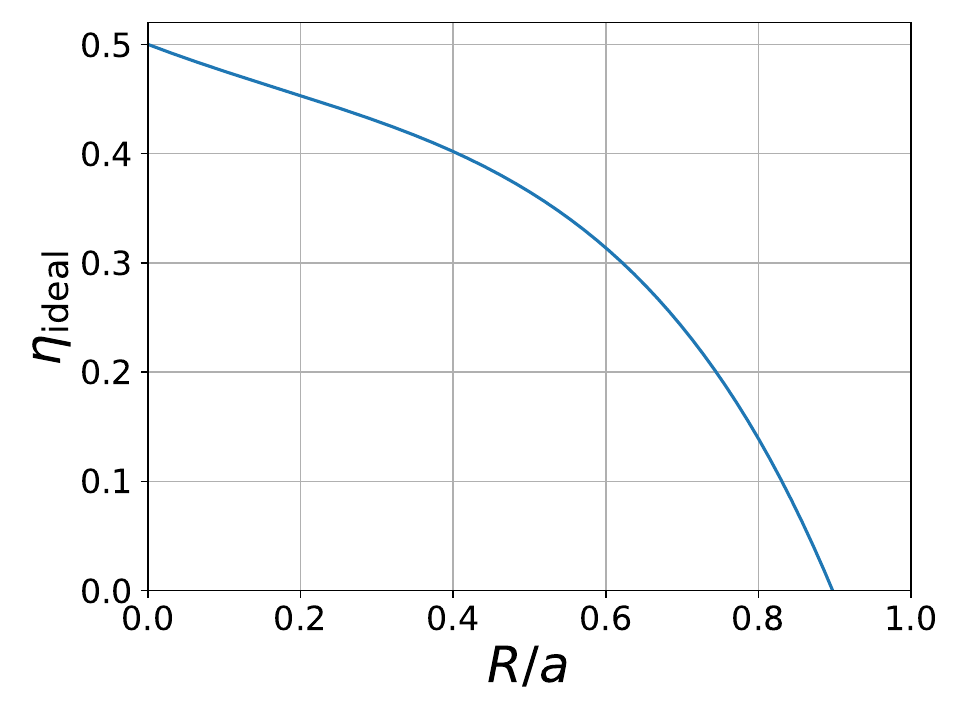}
    \caption{Ideal momentum transfer efficiency $\eta_{\rm ideal}$ versus $R/a$, based on (\ref{eq:etaideal}). For $R/a=0$, (\ref{eq:etaideal}) reproduces the result for planar ejecta $\eta_{\rm ideal}=1/2$.
    \label{fig:etaideal}}
\end{figure}

Consider a fluid parcel traveling outward from the supernova on a radial ray intersecting the spherical donor. The momentum imparted by the parcel when deflected through an angle $\sigma$ is
\be
p_{\rm imp} = p \cos\zeta\sin\sigma,
\ee
where $\zeta$ is the angle at which the parcel intersects the donor surface as measured from the center of the donor. These angles are illustrated in Fig.~\ref{fig:momentumdiagram} and are related by
\be
\sin\sigma = \frac{\cos\zeta-R/a}{\sqrt{1+(R/a)^2-2(R/a)\cos\zeta}}
\ee
for a donor with radius $R$ and orbital separation $a$. The radial ray tangent to the donor will occur when $\sigma=0$, corresponding to
\be
\zeta_{\rm max} = \pi/2-\arctan(R/a).
\ee
Thus we need only consider the portion of the donor surface with $\zeta\leq\zeta_{\rm max}$, as the remainder of the surface does not come into contact with ejecta in this formalism. The fraction of the total momentum of the ejecta intersecting the donor which is imparted to the donor is then
\be
\eta_{\rm ideal} = \frac{\int_{0}^{\zeta_{\rm max}} p_{\rm imp} \sin\zeta\cos\zeta d\zeta}{\int_{0}^{\zeta_{\rm max}} p \sin\zeta\cos\zeta d\zeta}.
\ee
Eliminating $\sigma$ and $p_{\rm imp}$, this may be written as
\be
\eta_{\rm ideal} = \frac{\int_{0}^{\zeta_{\rm max}} \sin\zeta\cos^{2}\zeta \frac{\cos\zeta-R/a}{\sqrt{1+(R/a)^2-2(R/a)\cos\zeta}} d\zeta}{\int_{0}^{\zeta_{\rm max}} \sin\zeta\cos\zeta d\zeta}. \label{eq:etaideal}
\ee
Note that as $R/a\rightarrow0$ this reduces to the case of planar ejecta
\be
\lim_{R/a\rightarrow0}\eta_{\rm ideal} = \frac{\int_{0}^{\pi/2} \sin\zeta\cos^{3}\zeta d\zeta}{\int_{0}^{\pi/2} \sin\zeta\cos\zeta d\zeta} = \frac{1}{2}.
\ee
This is also the maximum possible value of $\eta_{\rm ideal}$, as shown in Fig.~\ref{fig:etaideal}, suggesting that momentum transfer is less efficient for spherical ejecta. The efficiency decreases monotonically with $R/a$ but drops below zero only for clearly unphysical values of $R/a\gtrsim 0.9$. Because double-degenerate binaries have $0.3\lesssim R/a\lesssim 0.4$ at Roche lobe overflow, $\eta_{\rm ideal}$ generally sits just above 0.4.

To measure this efficiency in our Athena++ simulation we must first compute the initial momentum of the ejecta
\be
p_{\rm tot} = \int_{0}^{v_{\rm max}} 4\pi t^{3} v^{3} dv \rho(v).
\ee
Using the Gaussian ejecta profile presented in \citet{Wong_2024} and truncating material with $v>v_{\rm max}$, this evaluates to 
\be
\begin{aligned}
p_{\rm tot} = \frac{2}{\sqrt{\pi}} \Mej v_{0} \left[1+\left(1+\frac{v_{\rm max}^{2}}{v_{0}^{2}}\right)\right. \\
\left.\times\exp\left(-\frac{v_{\rm max}^{2}}{v_{0}^{2}}\right)\right].
\end{aligned}
\ee
Here $v_{0}=\sqrt{(4/3)\Eej/\Mej}$ is the characteristic velocity. The fraction of the ejecta which intersects the donor is
\be
\frac{\Omega}{4\pi} = \frac{1}{2} \int_{0}^{\arcsin(R/a)} \sin\theta d\theta.
\ee
We also need to project the momentum of this ejecta along the axis of symmetry, as transverse components cancel. This reduces the imparted momentum by a factor of 
\be
f_{\rm sym} = \frac{\int_{0}^{2\pi}\int_{0}^{\arcsin(R/a)}\sin\theta\cos\theta d\theta d\phi}{\int_{0}^{2\pi}\int_{0}^{\arcsin(R/a)}\sin\theta d\theta d\phi},
\ee
though this differs from unity by only a few percent. Then the total momentum of the ejecta imparted to the donor is $\Delta p = p_{\rm tot} f_{\rm sym} \Omega/4\pi = 187$ $\msun$km/s given the parameters of the explosion described in section \ref{sec:setup}. 

Our simulation results give a momentum transfer integrated over the donor surface of 96 $\msun$km/s, resulting in an efficiency of $\eta=96/187=0.513$. This is $\approx$20\% greater than the value predicted by (\ref{eq:etaideal}) of $\eta_{\rm ideal}=0.431$. This excess in $\eta/\eta_{\rm ideal}$ roughly matches the stiffest stellar models of \citet{2018ApJ...864..119H}, who found $\eta\approx 0.6$ for many of their $N=0$ and $N=1.5$ polytropes. However, our treatment of the donor as a rigid body eliminates the stellar response of the donor to the collision as a complicating factor. Thus, the fact that we also find deviations from theory illustrates the limitations of the assumption that each gas parcel follows a ballistic trajectory, neglecting collisions within the fluid.

\bibliography{references}{}
\bibliographystyle{aasjournalv7}


\end{document}